\documentclass[12pt,preprint]{aastex}
\usepackage{amsmath}
\usepackage{ccaption}

\newcommand{\Zsun}{\hbox{\it Z$_\odot$}}
\newcommand{\kms}{km~sec$^{-1}$}

\newcommand{\SiII}{Si\,\textsc{ii}}
\newcommand{\MgII}{Mg\,\textsc{ii}}
\newcommand{\FeII}{Fe\,\textsc{ii}}

\newcommand{\lam}[1]{\ensuremath{\lambda}\,#1}
\newcommand{\lamlam}[1]{\ensuremath{\lambda\lambda}\,#1}
\newcommand{\ca}{Ca\,\textsc{ii}~{\lam 3945}}
\newcommand{\csalt}{$c_{salt}$}
\newcommand{\dm}{$\Delta\!m_{15}$}
\newcommand{\fe}{\rm{Fe}\,\textsc{ii}~{\lam 4800}}
\newcommand{\mg} {Mg\,\textsc{ii}~{\lam 4300}}
\newcommand{\sia}{\rm{Si}\,\textsc{ii}~{\lam 4130}}
\newcommand{\sib}{\rm{Si}\,\textsc{ii}~{\lam 5972}}
\newcommand{\sic}{\rm{Si}\,\textsc{ii}~{\lam 6355}}
\newcommand{\sid}{\rm{Si}\,\textsc{ii}~{\lam 3858}}
\newcommand{\siia}{\rm{S}\,\textsc{ii}~{\lam 5454}}
\newcommand{\siib}{\rm{S}\,\textsc{ii}~{\lam 5640}}
\newcommand{\sw}{\rm{S}\,\textsc{ii}~W}
\newcommand{\uvb}{UV2}

\newcommand{\xx}{$x_1$}
\newcommand{\vel}[1]{#1 km~sec$^{-1}$}
\newcommand{\rchi}{$\chi^2$/d.o.f.}
\newcommand{\gold}{{\it gold}}
\newcommand{\silver}{{\it silver}}
\usepackage{color}
\usepackage{graphicx}
\usepackage{subfigure}

\shorttitle{line profiles of SNe~Ia}
\shortauthors{Konishi et al.}

\begin{document}

\title{Line Profiles of Intermediate Redshift Type~Ia Supernovae \altaffilmark{1}}

\author{Kohki Konishi\altaffilmark{2,3}, 
Joshua A. Frieman\altaffilmark{4,5,6}, 
Ariel Goobar\altaffilmark{7,8},
John Marriner\altaffilmark{4},
Jacob Nordin\altaffilmark{7,8},
Linda \"{O}stman\altaffilmark{9},
Masao Sako\altaffilmark{10},
Donald P. Schneider\altaffilmark{11},
Naoki Yasuda\altaffilmark{2,12}
}
\email{kohki@icrr.u-tokyo.ac.jp}

\altaffiltext{}{Department of Physics, Stockholm University, 106 91 Stockholm, Sweden.}
\altaffiltext{}{Oskar Klein Centre for Cosmo Particle Physics, AlbaNova, 106 91 Stockholm, Sweden.}

\altaffiltext{1}{Based in part on data collected at Subaru Telescope, which is operated by the National Astronomical Observatory of Japan.}
\altaffiltext{2}{Department of Physics, Graduate School of Science, University of Tokyo, Tokyo 113-0033, Japan.}
\altaffiltext{3}{Institute for Cosmic Ray Research, University of Tokyo, Kashiwa 277-8582, Japan.}
\altaffiltext{4}{Center for Particle Astrophysics, Fermi National Accelerator Laboratory, P.O. Box 500, Botavia, IL 60510, USA}
\altaffiltext{5}{Kavli Institute for Cosmological Physics, The University of Chicago, 5640 South Ellis Avenue Chicago, IL, 60637, USA}
\altaffiltext{6}{Department of Astronomy and Astrophysics, The University of Chicago, 5640 South Ellis Avenue Chicago, IL, 60637, USA}
\altaffiltext{7}{Department of Physics, Stockholm University, 106 91 Stockholm, Sweden.}
\altaffiltext{8}{Oskar Klein Centre for Cosmo Particle Physics, AlbaNova University Center, 106 91 Stockholm, Sweden.}
\altaffiltext{9}{Institut de F\'{i}sica d'Altes Energies, Universitat Aut\`{o}noma de Barcelona, Bellaterra, Spain.}
\altaffiltext{10}{Department of Physics and Astronomy, University of Pennsylvania, Philadelphia, PA 19104, USA.}
\altaffiltext{11}{Department of Astronomy and Astrophysics, Pennsylvania State University, University Park, PA 16802 USA.}
\altaffiltext{12}{Institute for the Physics and Mathematics of the Universe, University of Tokyo, Kashiwa 277-8582, Japan.}

\begin{abstract}
We present the temporal evolution of line profiles ranging 
from near ultraviolet to optical wavelengths by analyzing 59 Subaru 
telescope spectra of normal Type Ia Supernovae (SNe~Ia) in the intermediate 
redshift range ($0.05<z<0.4$) discovered by the Sloan Digital Sky 
Survey-II (SDSS-II) Supernova Survey.  We derive line velocities, peak 
wavelengths and pseudo-equivalent widths (pEWs) of these lines.  Additionally, 
we compare the line profiles around the date of maximum brightness with 
those from their nearby counterparts.  We find that line profiles 
represented by their velocities and pEWs for intermediate redshift SNe Ia are consistent 
with their nearby counterparts within $2\sigma$.  These findings 
support the picture that SNe Ia are a ``standard" candle for the intermediate redshift 
range as has been shown between SNe Ia at nearby and high redshifts.  
There is a hint that the ``{\mg}" pEW distribution for intermediate redshift 
SNe Ia is larger than for the nearby sample, which could be interpreted as 
a difference in the progenitor abundance.
\end{abstract}


\section{Introduction}
The diversity of type Ia supernovae (SNe Ia) has been extensively 
investigated over the past thirty years; see \citet{bra81,bra88,wel94} 
for initial studies.  Photometric lightcurves have revealed that the 
maximum luminosity is related to the decline rate of luminosity.  A 
parameter for SN lightcurve variations, {\dm}, is defined as the 
magnitude decline between the maximum brightness date and 15 days later.  
SNe Ia are found to be brighter with smaller values of {\dm} 
\citep[e.g.][]{phi99}.  The maximum luminosity is also related to (B-V) 
color \citep[e.g.][]{rie96,tri98}.  These relations play an essential 
role in the standardization of maximum luminosity and therefore the 
description of the accelerating universe \citep{rie98,per99}.  Spectra 
are also suited for the investigation of SN Ia diversity, since they 
can retain subtle information which would be blurred in photometry: 
chemical composition and excitation levels of ejected materials such as 
magnesium, silicon, sulfur, calcium, titanium, chromium and iron. 

The spectroscopic diversity of SNe Ia is seen in many features.  The 
large velocity scatter ($\sim 3,000$ {\kms}) of the location of the 
``{\sic}" line  has been presented in \citet{bra88}.  The temporal 
evolution of various optical lines is examined in \citet{fol04}: 
equivalent widths (pEWs) of the ``{\fe}"  feature evolve in a 
homogeneous way with phase, irrespective of their lightcurves.  The 
``{\mg}" pEW, however, increases abruptly at later times for SNe Ia 
with smaller {\dm}. The temporal evolution of ``{\sia}", ``{\sib}" 
and ``{\sic}" pEWs were also found to be heterogeneous.

The chemical composition of SN progenitors is supposed to have an important 
impact on the appearance of spectral features. \citet{len00} examined 
metallicity effects of progenitors on SN spectra at several phases with 
simulations, by scaling all elements heavier than oxygen in the unburned 
carbon/oxygen (CO) layer. UV features are shown to move blueward with 
higher metallicity. Since metallicity evolves with redshift \citep{rod08}, 
SN spectral features would be expected to change over cosmic time and 
perhaps bias the cosmological parameter measurements. \citet{coi00} 
examined two spectra of SNe Ia at high redshift (high-z) and reported no 
significant difference from nearby SNe Ia.
\citet{blo06} compared SN Ia spectra taken for the Equation of State: 
SupErNovae trace Cosmic Expansion survey \citep[ESSENCE;][]{mik07} with 
nearby SN Ia spectra in the SUpernova SPECTra(SUSPECT) database 
\footnote{\url{http://bruford.nhn.ou.edu/\~{}suspect/}}.
They measured line velocities of ``{\ca}", ``{\siia}", ``{\siib}" and 
``{\sic}" for nearby and high-z ($0.3\lesssim z \lesssim0.7$) SNe Ia 
to examine the temporal evolution of line velocities. This work, and 
that of \citet{gar07}, found no evolution between nearby and high-z SNe Ia.

A possible redshift-dependence in the temporal evolution of features has 
been reported recently.  Around half of the SN Ia spectra at high-z 
($z \sim 0.7$) taken for the SuperNova Legacy Survey \citep[SNLS;][]{ast06} 
show lower ``{\mg}" pEWs than the $1\sigma$ pEW distribution of their nearby 
counterparts at 5 to 10 days after the dates of maximum brightness 
\citep{bro08}.  The observed differences might be caused by selection effects 
in the high-z sample:  the abrupt increase in``{\mg}" occurs later 
in the brighter supernovae that tend to be selected for spectroscopic 
follow-up, especially at high-z.
A decrease of the ``{\fe}" pEW was reported by creating four composite 
spectra in the redshift range of 0.1 to 0.8.  The redshift evolution was 
further investigated by constructing composite spectra of nearby, SNLS/Keck, 
and Hubble Space Telescope (HST) reference samples \citep{sul09}.
They claimed a lower intermediate mass element (IME) abundance toward 
$z \lesssim 1$ by showing the drop of average pEW for ``{\ca}", ``{\sia}", 
and ``{\mg}". 
The extent of SN Ia diversity and their possible evolution is still uncertain.
While SNe Ia spectra in the nearby and high-z ranges have often been 
obtained, few SNe Ia spectra in the intermediate redshift range have been observed. 

The Sloan Digital Sky Survey-II (SDSS-II) Supernova Survey 
\citep{fri08,sak08} has obtained over 500 spectroscopically confirmed 
SNe Ia in the intermediate redshift range.  Lightcurves and spectra of the first year 
sample are described in \citet{hol08} and \citet{zhe08}.
A subset of SDSS-II SN Ia spectra by the New Technology Telescope and the 
Nordic Optical Telescope (NTT/NOT) was used for an optical spectral 
feature study \citep{ost10,nor10}. 
We investigate near UV to optical line profiles of SNe Ia obtained by a 
different reduction pipeline, galaxy subtraction method, and selection 
criteria \citep{kon09a} by comparing a sample of nearby SN Ia spectra with 
those obtained by the Subaru telescope in the course of the SDSS SN survey.  
The following is the outline of this paper. 
The data are summarized in \S \ref{data09b}.
\S \ref{measure09b} gives the measurement methods of spectral features. 
The temporal and redshift evolution of these features are presented in 
\S \ref{result}.
We discuss and conclude this paper in \S \ref{disc09b} and \S \ref{sum09b}. 

\section{Data} \label{data09b}
\subsection{Intermediate redshift Supernovae}
An extension of the Sloan Digital Sky Survey \citep{yor00}, the SDSS-II Supernova 
Survey, obtained lightcurves of hundreds of SNe~Ia at the intermediate redshift, 
$0.05 < z < 0.40$ \citep{fri08,sak08,hol08}.
The SN survey have acquired observations from September to November during 
2005 - 2007 on the 300 square degrees of the equatorial region 
\citep[Stripe 82;][]{sto02}.
The SDSS 2.5m telescope \citep{gun06} and camera \citep{gun98} have been 
used with $ugriz$ filters \citep{fuk96} to repeatedly scan the wide field of sky.
Spectroscopy for SN identification and redshift determination was performed 
with various telescopes including the Subaru telescope \citep{kon09a}; see 
also \citet{sak08,zhe08,ost10,fol10}.

We observed 62 SN~Ia spectra with the Faint Object Camera and Spectrograph 
(FOCAS; \citet{kas02}) mounted on the Subaru telescope during 2005 and 2006.
The instrumental wavelength range was 3600 {\AA} (4000 {\AA} for Year 2005) 
to 6000 {\AA} for blue side spectra and 5000 {\AA} to 9000 {\AA} for red 
side spectra at a spectral resolution of 500.  We generally took blue and 
then red side spectra for each SN and combined them using flux densities 
at overlapping wavelengths.
SN observations were not typically made at the parallactic angle, but the atmospheric refraction was corrected by an atmospheric dispersion corrector.
Phases of SN Ia spectra, elapsed rest-frame days from the maximum date 
($p$), are determined by the SALT2 lightcurve fitter \citep{guy07}.  
Lightcurve parameters of lightcurve width {\xx} and color {\csalt} are 
also determined \citep{kon09a}.  Each of intermediate redshift SNe Ia sample 
is determined as {\gold} or {\silver} SNe using the following photometric and 
spectroscopic criteria.

{\bf Lightcurves: }
Since the season for photometric observations was limited, some 
lightcurves do not have sufficient phase coverage to accurately determine 
photometric properties. We set the following criteria for their lightcurves:
\begin{enumerate}
 \item at least one data point with $p<-4$.
 \item at least one data point with $p>+4$. 
 \item at least five data points with $-20<p<+60$.
 \item lightcurve parameters with $|x_1|<5.0$ and/or $|c_{salt}|<2.0$, 
which are the SALT2 limits of these parameters.
\end{enumerate}
The criteria for lightcurves are similar to those for the cosmology sample 
\citep{kes09} except that our requirement for the date of first observation 
is slightly more stringent to securely determine the dates of maximum luminosity.

{\bf Spectra: }
The observed SN spectra generally also contained light from host galaxy.  
In order to examine line velocities and pEWs precisely, SN spectra were 
extracted from 2 Dimensional (2D) galaxy-contaminated spectra by fitting 
each spatial profile with point and broad components without the aid 
of galaxy template spectra \citep{kon09a}.  This method is similar to that 
of \citet{blo05,bau08}.  The 2D extraction method worked well for most of 
our intermediate redshift spectra but failed when the bestfit width for a SN component 
was incompatible with the seeing size, which can arise if a SN was located 
too near the center of its host galaxy. 
We set the following criteria for spectral extraction: 
\begin{enumerate}
 \item the separation of the SN and the host center was larger than 1 pixel 
(0.312 arcsec) or there was no apparent host
 \item a bimodal spatial profile was found: a minimum appeared in the 
signal between the SN and its host center.
\end{enumerate}

\begin{figure}
 \plotone{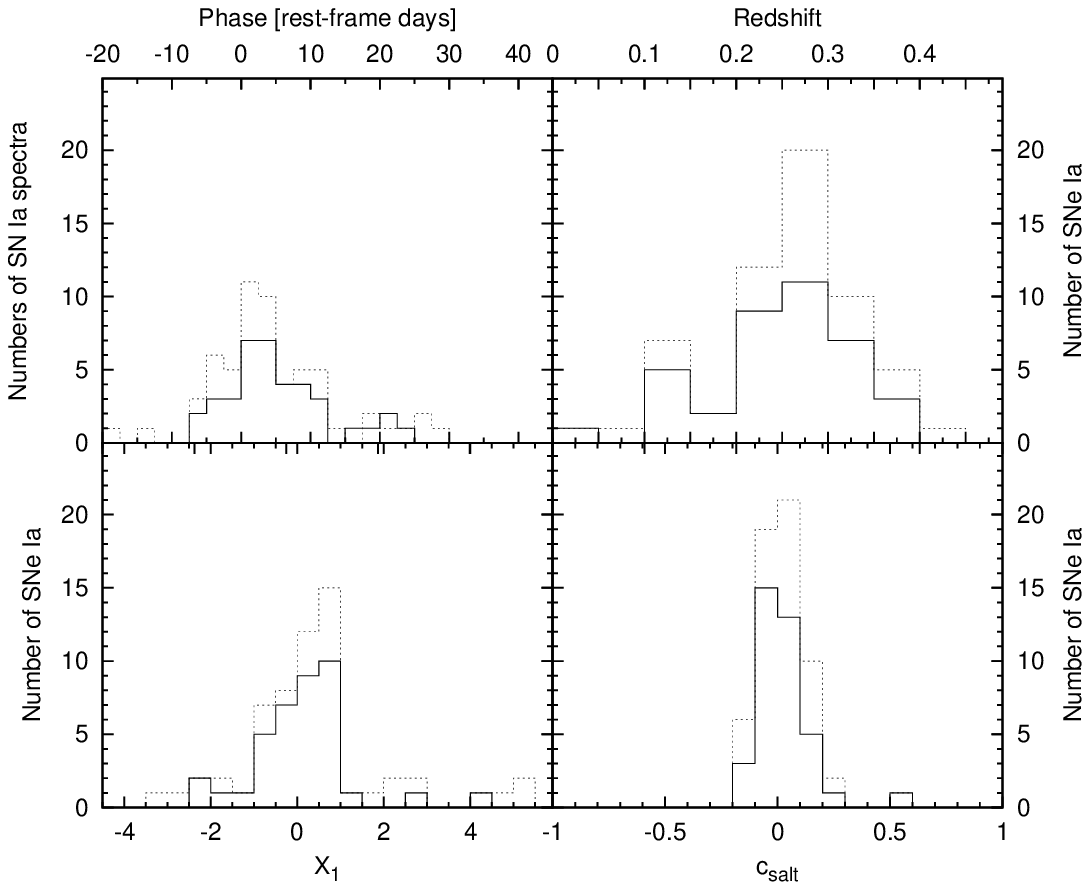}
 \caption{Distributions of SN phase (top left), redshift (top right), lightcurve width {\xx} (bottom left) and color {\csalt} (bottom right) for the intermediate redshift sample; total of 59 SNe Ia (dotted) and 38 {\gold} SNe Ia (solid).
 \label{sbrdist}}
\end{figure}

Table \ref{sbr1aspecinfo} summarizes normal intermediate redshift SNe Ia observed by the 
Subaru telescope.  The SDSS internal identification number (ID) and the 
IAU name are written in Columns 1 and 2.  The Galaxy extinction from 
\citet{sch98} is listed in Column 3.  The time of each observation is 
in Column 4.  The redshift and the method to obtain the redshift are shown 
in Column 5: ``ge" indicates that the redshift of the target was determined 
from galaxy emission line(s), ``ga" from galaxy absorption line(s) and 
``sn" from fitting the SN spectrum.  Fore more details about the redshift 
determination see \citet{kon09a}.  Column 6 is the validity of the data. 
We regard those SNe Ia which passed all of the lightcurve and spectra criteria as 
the {\gold} sample, which is indicated as ``G" in Column 6. 
If a SN Ia fails at least one of the lightcurve (spectral) criteria, it is 
regarded as photometrically {\silver} (spectroscopically {\silver}), which is 
indicated as ``PS (SS)" {\silver}. Those which failed both photometric and 
spectroscopic criteria are indicated as ``S" in Column 6.
A total of 38 SNe Ia are regarded as {\gold} and 21 as {\silver}.  Of the 21 
{\silver} sample, 10 are photometrically classified as {\silver}, another 
9 are spectroscopically classified, and the other 2 are both.  SNe Ia 
classified as {\silver} by the fourth lightcurve criterion are SN2006jf and 
SN2006jv and those by the second spectral criterion are SN2005hz, 2006qa 
and 2006qe.  Column 7 contains the phase of the SN Ia spectra.
Only one SN Ia spectrum was observed for each object in our sample.
The distributions of phase, redshift, lightcurve width {\xx}, and color 
{\csalt} for 59 intermediate redshift SNe Ia (dotted) and 38 {\gold} SNe Ia (solid) are 
shown in Figure \ref{sbrdist}.  The Subaru sample is smaller (59 versus 141 SNe Ia)
than the NTT/NOT sample \citep{ost10,nor10} but higher redshift ($0.25 \pm 0.08$ versus 
$0.17 \pm 0.07$).

\subsection{Control sample}
We use two types of nearby normal SNe~Ia datasets for comparison of 
line features with the intermediate redshift sample. 
Here we call SNe Ia except for 1991bg-like, 1991T-like and peculiar (in any sense) 
objects as normal ones.
One consists of optical spectra ({\bf Opt}) and the other of UV spectra ({\bf UV}).

\begin{figure}
 \plotone{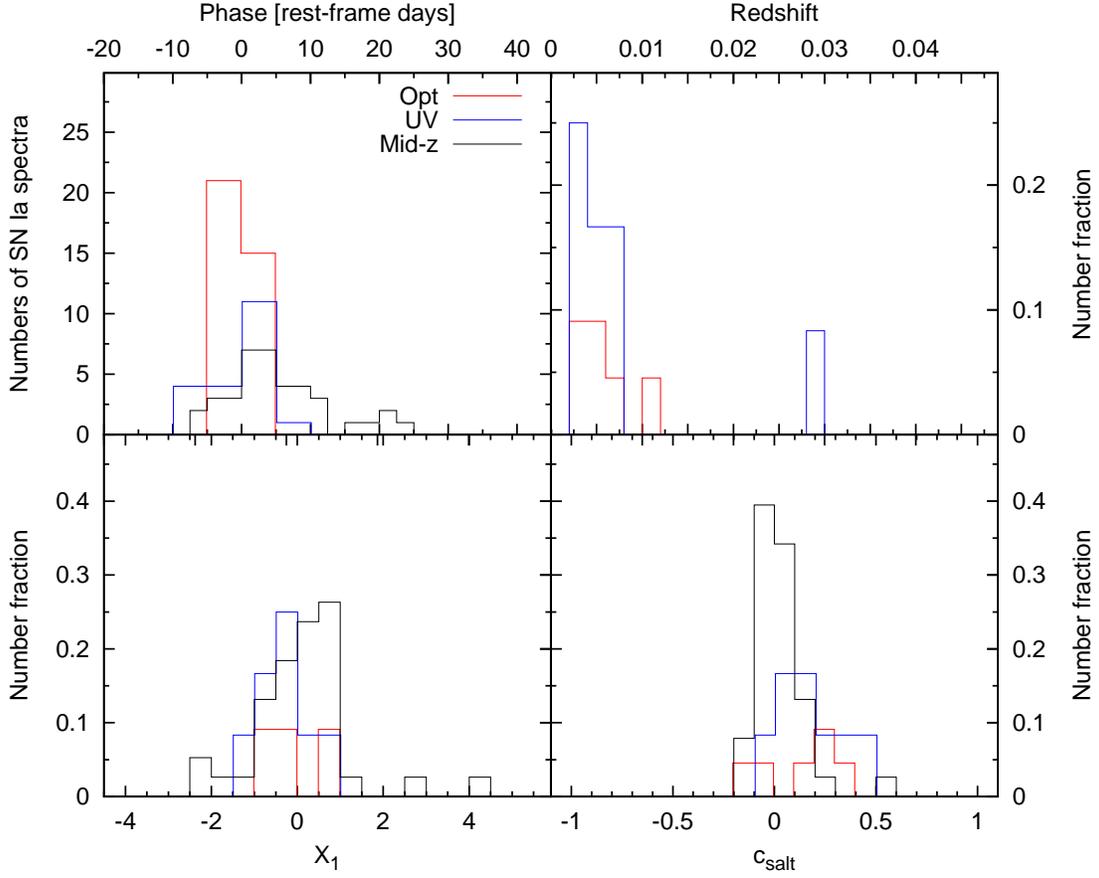}
 \caption{Distributions of SN phase (top left), redshift (top right), lightcurve width {\xx} (bottom left) and color {\csalt} (bottom right) for two control samples, the {\bf  Opt} (red) and {\bf UV} (blue) and 38 intermediate redshift {\gold} sample (black) for a reference. The histograms for redshift, lightcurve width {\xx} and color{\csalt} are normalized so that the area under the intermediate redshift sample is the same as for the reference samples.
 \label{sampledist}}
\end{figure}

The {\bf Opt} sample consists of 121 spectra at $-10<p<+40$ from 12 
well-observed SNe~Ia.  This sample is the same sample examined by 
\citet{bro08} except that the sample lacks SN1991T-like, SN1991bg-like 
and unpublished samples.  We downloaded spectra from the SUSPECT database 
and removed the atmospheric A line (absorption at 7600 {\AA}) from them 
if we found it near 7600 {\AA} in their observed frame. 
The {\bf UV} sample consists of 20 spectra at $-10<p<+10$ from 8 SNe Ia.  
We use the public spectral dataset obtained with the International 
Ultraviolet Explorer and the HST \citep{fol08b} and chose spectra in the  
wavelength region from 3050 to 3250 {\AA} from the dataset.  
SN1991bg-like or SN1991T-like samples were not included.  
Rest frame $BVR$-band lightcurves are comparable to $gri$ bands for the 
intermediate redshift SNe Ia. 
Lightcurves were gathered from various papers \citep{but83,cia88,but85,phi87,ham91,for93,lei93,wel94,lir98,jha99,rie99,kri01,kri03,kri04,val03,kri04,alt04,anu05,rie05}.
A $B$-band lightcurve for SN1991M and $R$-band lightcurves for SN1982B, SN1986G and SN2001ba were missing.
The maximum date and other lightcurve parameters for the {\bf Opt} and {\bf UV} samples were obtained using the SALT2 \citep{guy07} code as for the intermediate redshift sample.
Figure \ref{sampledist} shows distributions of spectral phases (upper left), redshift (upper right), lightcurve width {\xx} (lower left) and color {\csalt} (lower right) for the nearby sample of {\bf Opt} (red) and {\bf UV} (blue) together with the intermediate redshift {\gold} SNe Ia (black). 

Tables \ref{optspecinfo} and \ref{uvspecinfo} summarize the details of 
the {\bf Opt} and {\bf UV} spectra, respectively.  Column 1 is the IAU 
name for each SN.  SN coordinates are shown in Columns 2 and 3.  Column 
4 is the Galactic color excess $E(B-V)$ from the \citet{sch98} reddening 
map and Column 5 is the redshift.  Redshifts were corrected using the NASA/IPAC 
Extragalactic Database (NED) \footnote{\url{http://nedwww.ipac.caltech.edu/}}. 
The Galactic extinction was corrected using the \citet{car89} law and $E(B-V)$ values.

Additionally, we used the SN Ia spectral template of \citet{hsi07}.  This 
template was constructed from a stack of observed spectra.  Since this 
template covers the UV-to-optical wavelength region during $-10<p<+40$, 
we used it to estimate the mean and dispersion of nearby and intermediate redshift SNe Ia 
(\S \ref{evo}).

\section{Measurements} \label{measure09b}
\begin{figure}
 \plotone{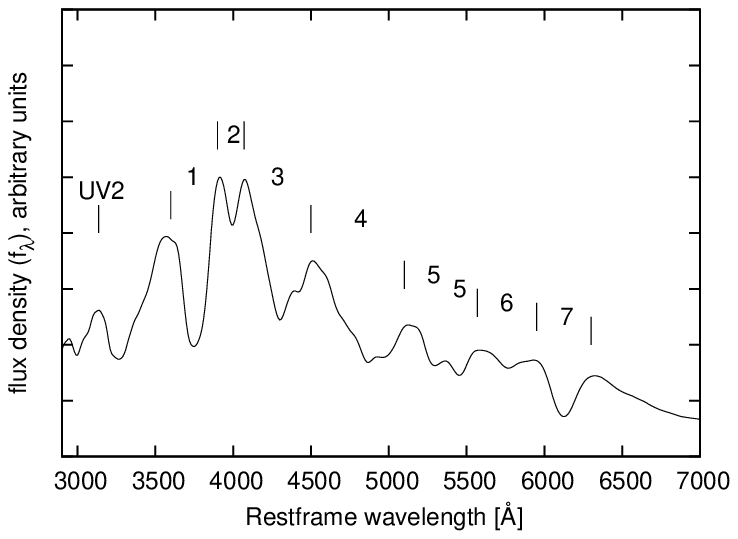} 
 \caption{SN Ia spectral feature definitions at near maximum brightness:
 Labels correspond to the following feature names:
 UV2- "{\uvb}", 1- "{\ca}", 2- "{\sia}", 3- "{\mg}", 4- "{\fe}",
 5- "{\siia}" and "{\siib}", 6- "{\sib}" and 7- "{\sic}". See also
 Table \ref{feature}. 
 The wavelengths of absorption lines are shifted blueward due to the expansion
 of SN ejecta.  Differences of absorption minima from their nominal wavelengths 
 (line velocities) have been measured as described in \S \ref{velmeasure}.
 \label{sn1aspec_b}}
\end{figure}

The definition of \citet{fol04} for each absorption feature is adopted 
(Figure \ref{sn1aspec_b}).  Note that this convention can cause some 
confusion.  Some observed lines (e.g. ``{\sia}") are a mixture of several 
lines from different elements, which can vary with time, temperature, and 
abundances \citep{bra08}.  We measured line velocities and pEWs for 
``{\ca}" (feature 1), ``{\mg}" (feature 3), ``{\fe}" (feature 4), and 
``{\sic}" (feature 7) from spectra in the range $-10<p<+40$, but  
we only measured these quantities for ``{\sia}" (feature 2), 
``{\SiII}{\lamlam 5454,5640}" (feature 5) and ``{\sib}" (feature 6) 
from spectra in the range $-10<p<+10$ when those features are most pronounced.
We also measured the wavelength for the ``{\uvb}" feature, locating between the 
absorption lines of {\MgII} and {\FeII} at $\sim 3200$ {\AA} \citep[e.g.][]{wei03}.

\subsection{Spectrum smoothing}
Since line forming regions are expanding, spectral features are 
intrinsically broadened and noisy spectra can be smoothed in order that as
little intrinsic information is missed as possible.
We smoothed intermediate redshift spectra $f_{obs}(\lambda)$ by using their uncertainty 
$f_{err}(\lambda)$: 
each point in the spectrum is weighted according to the value of its 
inverse variance.  A smoothed spectrum $\bar{f}(\lambda)$ is obtained as
\begin{equation}
 \bar{f}(\lambda) 
  = \frac{ \sum_{\lambda'} W(\lambda, \lambda') f_{obs}(\lambda') }
  { \sum_{\lambda'} W(\lambda, \lambda') },
\end{equation}
where the weight $W(\lambda, \lambda')$ is
\begin{eqnarray}
 W(\lambda, \lambda') 
  &=& \frac{G(\lambda, \lambda', \sigma_G)}{f_{err}(\lambda')} \\
 G(\lambda, \lambda', \sigma_G) &=& \frac{1}{\sqrt{2 \pi}} 
  \exp \Bigl( -\Bigl( \frac{\lambda-\lambda'}{\sigma_G} \Bigr)^2 \Bigr).
\end{eqnarray}
This procedure given in \citet{blo06} is based on the intrinsic broadening of 
SN features due to large expansion velocities their in line forming regions, and 
takes into account flux uncertainties which depend on wavelengths. 
Assuming the velocity of this broadening to be $1500-3000$ {\kms}, one can 
write a smoothing factor $\sigma_G = (v/c) \lambda$. 
We set $v=1500$ {\kms} for optical wavelengths and $v=3000$ {\kms} for UV 
wavelengths. This relatively larger velocity avoids misidentification 
of the ``{\uvb}" peak wavelength from several spikess which 
probably result from high-frequency noises.

\begin{figure}
 \plotone{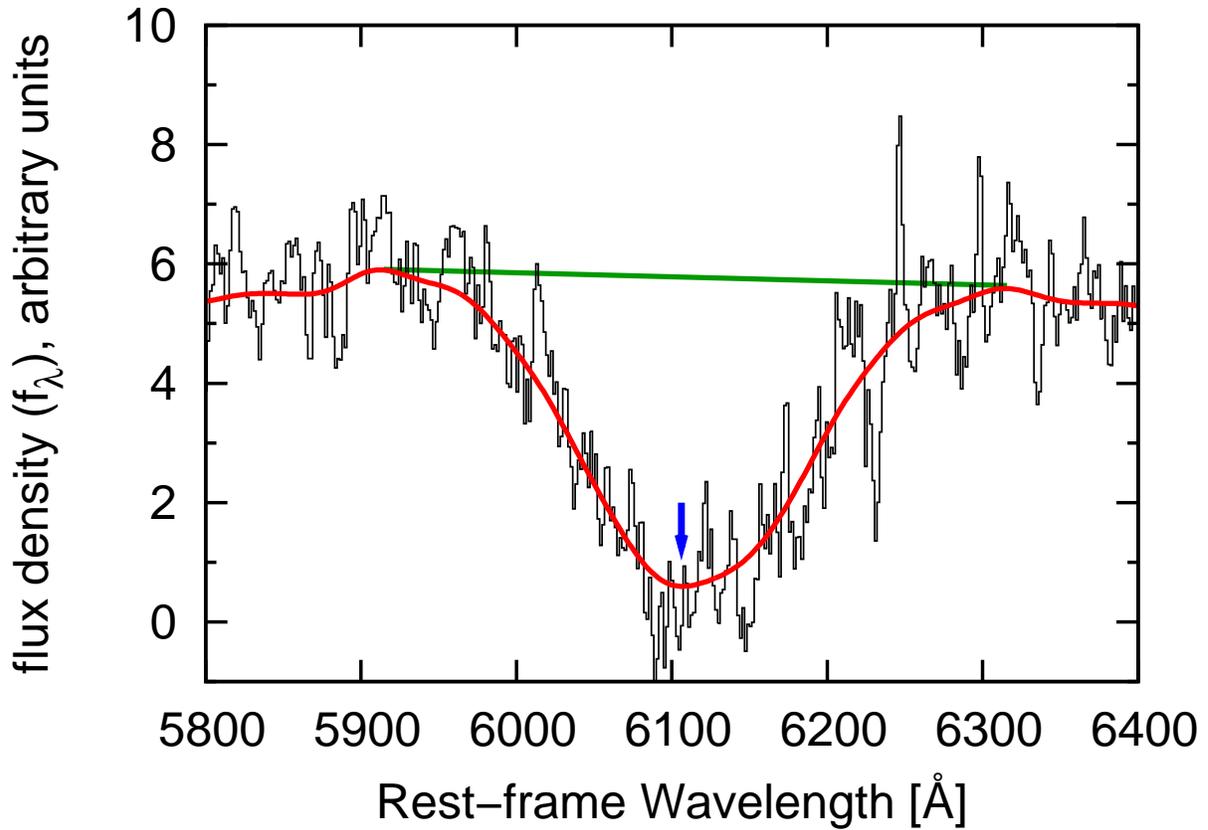}
 \caption{An example of velocity and pEW measurements. 
 The black histogram is the un-smoothed Subaru spectrum taken near the maximum brightness (SN16776). The smoothed spectrum is shown in the red curve. The pseudo continuum is determined for the "{\sic}" feature (green line).
 The absorption minimum wavelength (vertical arrow) in the smoothed spectrum is used to determine an absorption minimum and absorption boundaries for an pEW measurement. 
 \label{lineex}}
\end{figure}

Figure \ref{lineex} is an example of the result of this convolution.  
The black line is the observed spectrum, and the red line the smoothed 
spectrum.  Since absorption features are determined more clearly in a smoothed 
spectrum than in its noisy spectrum, we use smoothed spectra for the measurements
of line velocities and pEW.

\subsection{Line velocity measurements \label{velmeasure}}
A line velocity $v_{abs}$ is determined by using the relativistic Doppler formula.
\begin{align}
v_{abs} = \frac{ (\lambda_{obs} / \lambda_{rest})^2 - 1}
{ (\lambda_{obs} / \lambda_{rest})^2 + 1} c, \label{vabs}
\end{align}
where $\lambda_{obs}$ is the wavelength of an absorption minimum. 
The smoothed flux at the wavelength takes its minimum 
within the wavelength region for the feature.
The vertical arrow in Figure \ref{lineex} is an example of the absorption
minimum. 
The wavelength regions and representative wavelengths of all the features 
are summarized in Columns 3 and 5 of Table \ref{feature}.
Although a ``{\ca}" feature sometimes shows a 'W' shape due to the 
additional contribution of the ``{\sid}" line, we assumed that the 
largest contribution to this absorption is ``{\ca}", and measured the 
wavelengths of the deepest absorption minimum for ``{\ca}". 

Since an absorption minimum of a P Cygni profile is shifted blueward, 
absorption line velocities are negative. Following \citet{blo06}, 
we use the words {\it increase} and {\it decrease} literally, e.g. 
the velocity increases from \vel{-10,000} to \vel{-9,000}, and 
decreases from \vel{-10,000} to \vel{-15,000}.

\subsection{Pseudo-equivalent width measurements}
A ``pseudo" EW (pEW) of an absorption feature is defined to be 
an EW for a pseudo-continuum \citep{gar07,bro08,fol08a,sul09}. 
The pseudo-continuum is often assumed to change linearly, as shown 
in the green line in Figure \ref{lineex}, 
between two spectral peaks of the features next to 
the absorption (we call these peaks ``absorption boundaries").
The pEW is determined as follows:
\begin{align}
 pEW &= \sum_{i_{w}=1}^N \left( 1-\frac{ f_{obs}(\lambda_{i_{w}}) }
 { f_c(\lambda_{i_{w}})} \right) \Delta \lambda_{i_{w}},
\end{align}
where $N$ is the number of wavelength pixels underneath the 
pseudo-continuum $f_c$. 
The wavelengths of the two absorption boundaries ($\lambda_1$ and 
$\lambda_N$) are determined from the smoothed spectrum $f_{obs}$ 
by searching for peak fluxes within the wavelength regions in Column 4 
of Table \ref{feature}.
$\Delta \lambda_{i_{w}}$ is the wavelength interval between 
neighboring wavelengths.

{\bf Uncertainty estimation for line velocities and pEWs: }
We measure wavelengths of absorption minima, a bump peak and absorption 
boundaries from smoothed spectra.  Although the intrinsic broadenings of 
3,000 {\kms} or 1,500 {\kms} are typical, this process might bias the 
wavelengths. 
We estimate the uncertainty in the line measurements using a Monte Carlo 
simulation. There is a ``blank" region in 2D spectra which was far from 
the SN and far from the galaxy, so that only background light was reaching 
the detector.
We compute the flux error $f_{err}(\lambda)$ from the sky background of 
the actual spectrum added in quadrature to the photon counting errors of 
the smoothed spectrum.  The simulated spectrum is the actual smoothed 
spectrum smeared by the computed error.
This procedure is repeated 100 times to derive the average and standard 
deviation of absorption minimum wavelength or pEWs. Uncertainties on the 
line velocities are set to the velocity corresponding to the standard 
deviation of the measured absorption minima.
The statistical errors for the {\bf Opt} sample are negligible.  We use 
a typical galaxy rotational speed (\vel{200}) for the velocity error 
and assign no error for the pEW measurements.

\section{Spectral features} \label{result}
\subsection{Temporal evolution \label{phase}}

The signal-to-noise ratio (S/N) range for most of the intermediate redshift SN Ia spectra 
is from 3 to 10 per 2 {\AA} \citep{kon09a}.  Typical SN Ia features are 
easy to recognize by visual inspection.  We show the temporal evolution 
measurements of line velocities and pEWs.

{\bf Line velocities: }
\citet{blo06} measured line velocities of ``{\ca}", ``{\siia}", ``{\siib}" 
and ``{\sic}" for 36 high-z SNe Ia discovered by the ESSENCE survey. 
The temporal evolution of line velocities for the high-z sample was reported to 
be consistent with those for the nearby sample.
\citet{ell08} measured line velocities of ``{\uvb} and ``{\sia}" for 36 
high-z SNe Ia from the SNLS/Keck sample to examine the behavior of UV 
features compared to an optical line.

\begin{figure}
\plottwo{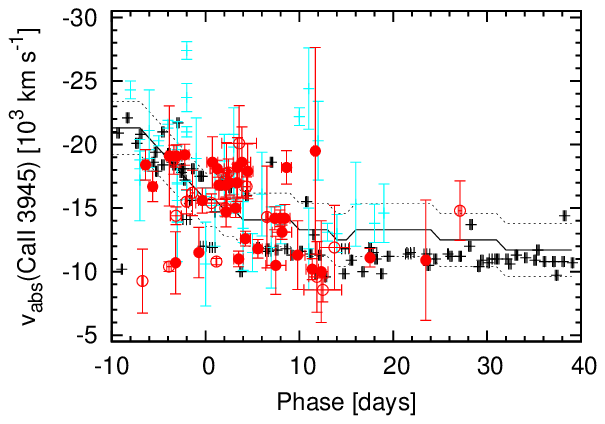}
{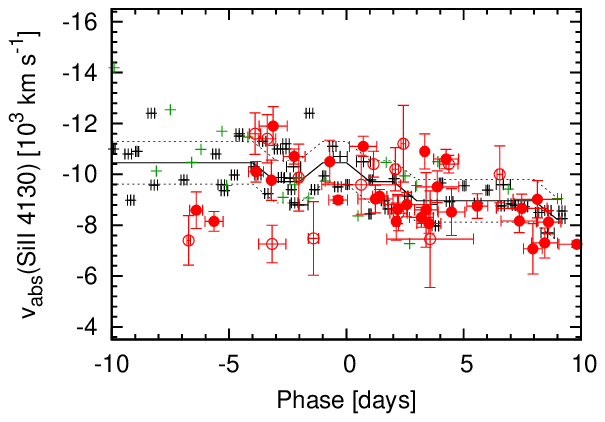} \\
\plottwo{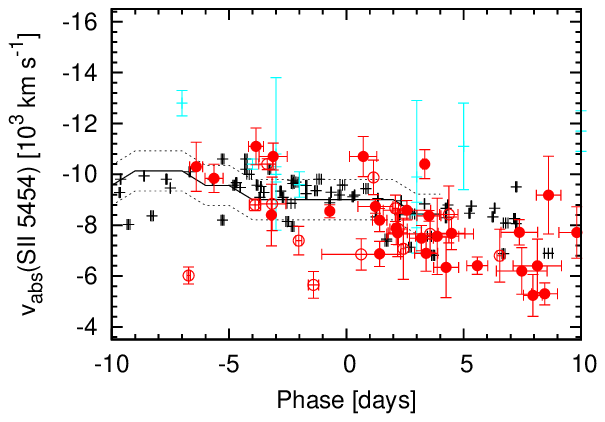}
{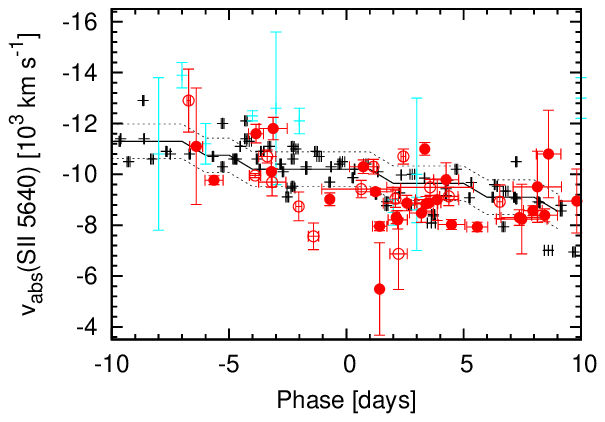} \\
\plottwo{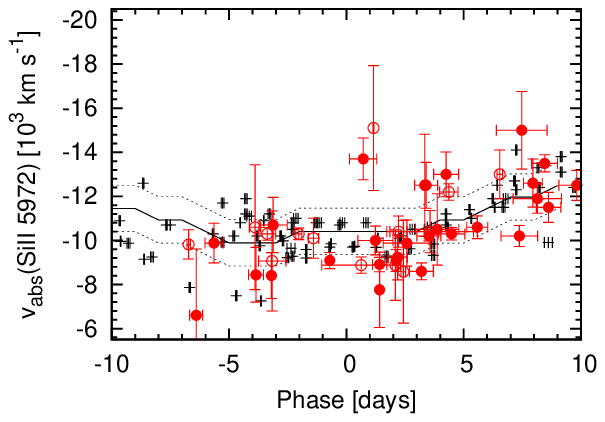}
{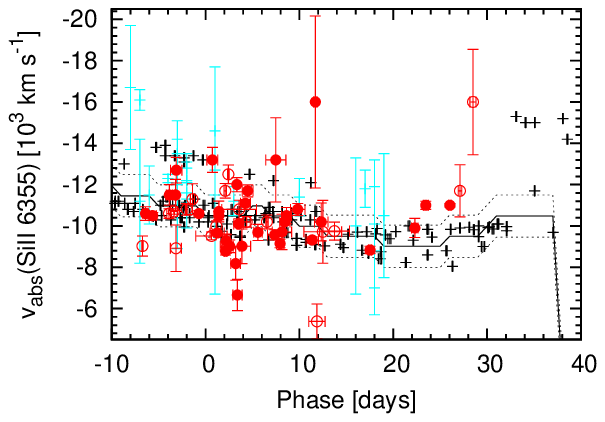} \\
 \caption{The temporal evolution of absorption line velocities for various 
 features; phases are rest-frame days.  The intermediate redshift sample is shown as red 
 circles (filled circles for the {\gold} sample and open circles for the 
 {\silver} sample.  Details are described in \S \ref{phase}).  The {\bf Opt} 
 sample is given as black pluses.  Overplotted are the results of 
 \citet{blo06} (cyan pluses) and \citet{ell08} (green pluses).  The 
 measurements for the \citet{hsi07} template shifted to represent the 
 nearby mean trend are shown as the solid line. Two dotted lines 
 are $\pm 1\sigma$ of the intermediate redshift sample.
  \label{p_vel}
}
\end{figure}

In Figure \ref{p_vel}, we plot the temporal evolution of line 
velocities for six features of nearby and intermediate redshift SNe Ia: (a) ``{\ca}", 
(b) ``{\sia}", (c) ``{\siia}", (d) ``{\siib}", (e) ``{\sib}", and 
(f) ``{\sic}".  intermediate redshift SNe Ia are shown as circles (filled for {\gold} 
SNe Ia and open for {\silver} SNe Ia) and the nearby sample as black 
pluses. 
As has been found previously in nearby and high-z SNe Ia 
\citep[e.g.][]{blo06}, one can see the following for the intermediate redshift SNe Ia: 
(i) the range of line velocities for the ``{\ca}" line is wide, from 
\vel{-20,000} to \vel{-10,000} at $-7 \lesssim p \lesssim 0$.
This range converges to \vel{a few $\times 10^3$} at $p \sim 30$.
(ii) The ``{\sia}" line velocity is almost constant at \vel{-11,000} 
for $-7 \lesssim p \lesssim 0$ and smoothly increases by 
\vel{$\sim$ 3,000} to $p=5$. 
(iii) The velocity for the ``{\siia}" line is \vel{-10,000} at  
$-7 \lesssim p \lesssim -3$ and increases only gradually to \vel{-7,000} 
at  $-7 \lesssim p \lesssim -3$. 
(iv) The ``{\siib}" line velocity is \vel{-11,000} at 
$-7 \lesssim p \lesssim -3$ and gradually increases to \vel{-9,000} afterwards.
(v) The ``{\siia}" velocity is always larger than the ``{\siib}" velocity at a fixed phase. 
\citet{bra08} explains this increase as the contamination of iron peak 
elements, and the ``{\siia}" line becomes undetectable at $p \sim 10$.
(vi) The ``{\sib}" line velocity is constant around \vel{-10,000} at  
$-7 \lesssim p \lesssim +3$ and slightly decreases afterwards.  
\citet{bra05} explains this increase as the strengthening of the NaI D line. 
(vii) The ``{\sic}" line velocity is constant around \vel{-11,000} for 
$-7 \lesssim p \lesssim -3$ and increases by a few \vel{$\times 10^3$} 
until $p \sim 10$.  Several measurements of SN1998bu show large negative 
velocities at $30<p<40$.  \citet{bra08} explains the velocity decrease at 
$30 \lesssim p$ as the inclusion of {\FeII}.
Typical velocities for Ca, Si and S are \vel{-15,000}, \vel{-10,000} and 
\vel{-9,000} at maximum brightness (Figure \ref{p_vel}), which confirms
an onion-like structure of the SN interior
\footnote{Elements with large negative velocity are distributed in outer 
layers.}.

\begin{figure}
 \plotone{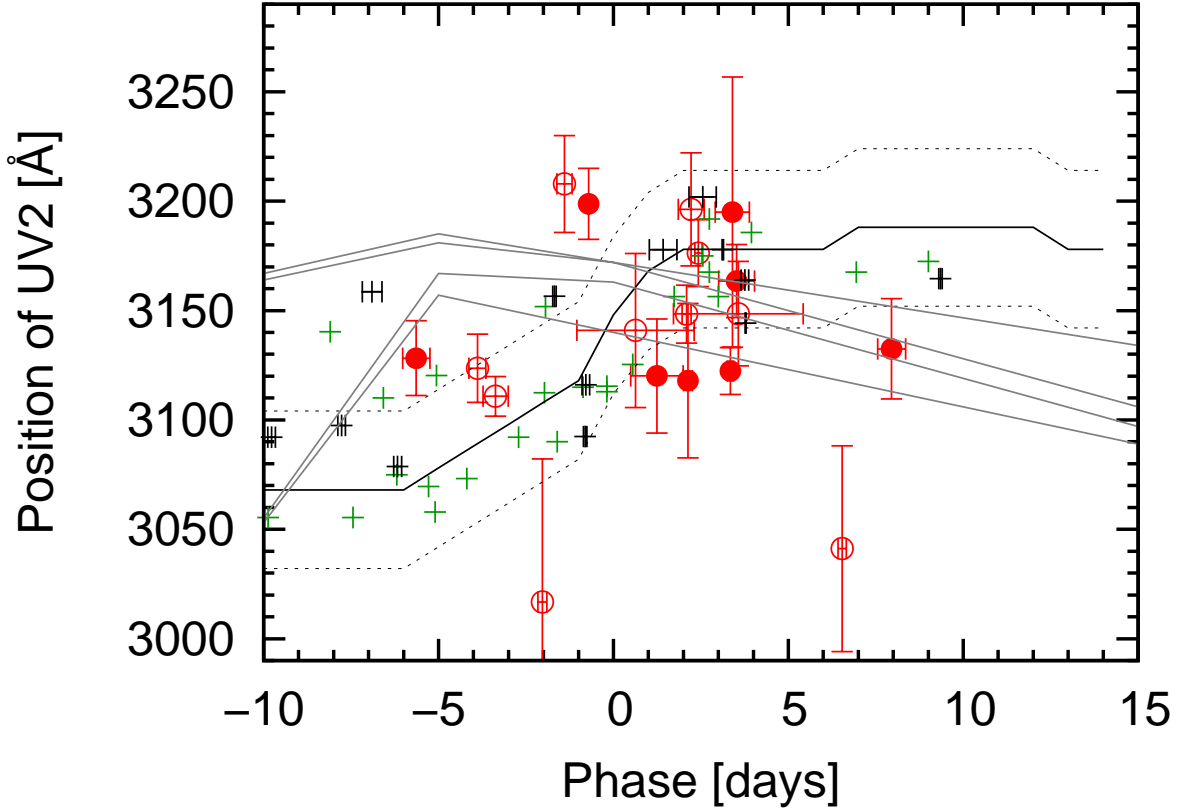}
 \caption{Peak wavelengths for the "{\uvb}" feature plotted versus rest-frame phase for the Subaru data (red circles) and the {\bf UV} sample (black pluses).  
 Overplotted are the results of \citet{ell08} (green pluses).
 The four gray lines represent the results of theoretical templates of \citet{len00} for the different metallicities in the unburned CO layer (top to bottom: 0.1, 0.3, 1.0 and 3.0 times solar metallicity).  
 The measurements for the \citet{hsi07} template shifted to represent the nearby mean trend are shown as the solid black line. Two dotted lines are $\pm 1\sigma$ of the intermediate redshift sample.
 \label{p_wvuv2}
 }
\end{figure}

In Figure \ref{p_wvuv2}, we show the peak wavelength of the ``{\uvb}" 
bump against SN phases for the intermediate redshift and {\bf UV} sample (the same labels 
as in Figure \ref{p_vel}).  The ``{\uvb}" wavelength for intermediate redshift {\gold} 
SNe Ia ranges from 3100 to 3200 {\AA} throughout the range of phases of 
the observations. 

{\bf Pseudo-equivalent widths: }
\citet{gar07} measured ``{\ca}", ``{\mg}" and ``{\fe}" pEWs for 12 SNe Ia 
($0.2<z<0.9$) discovered by the SCP and reported a consistency between 
the temporal evolutions of pEWs for nearby and high-z SNe Ia.
\citet{bro08} measured ``{\ca}", ``{\sia}" and ``{\mg}" for 46 SNLS/Gemini 
SNe Ia and found a marginal difference only for the ``{\mg}" feature.
This study contains more features than these previous studies.

\begin{figure}
\plottwo{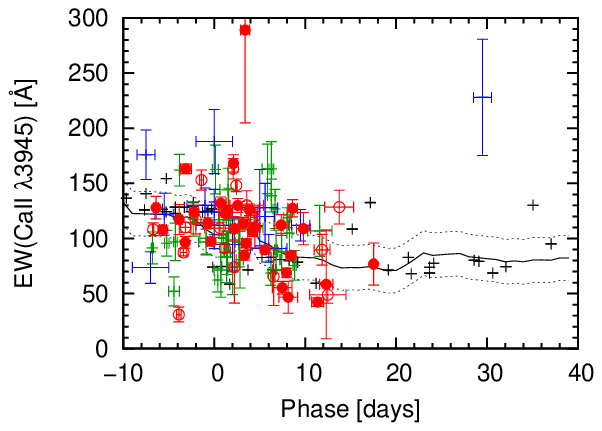}
{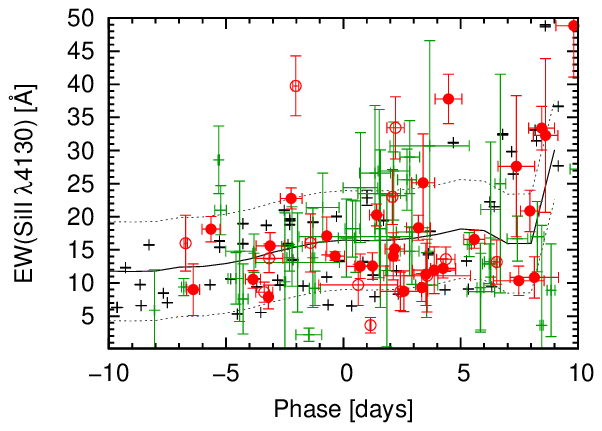} \\
\plottwo{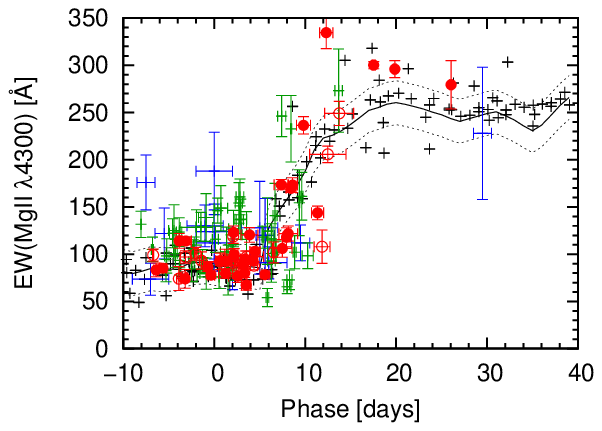}
{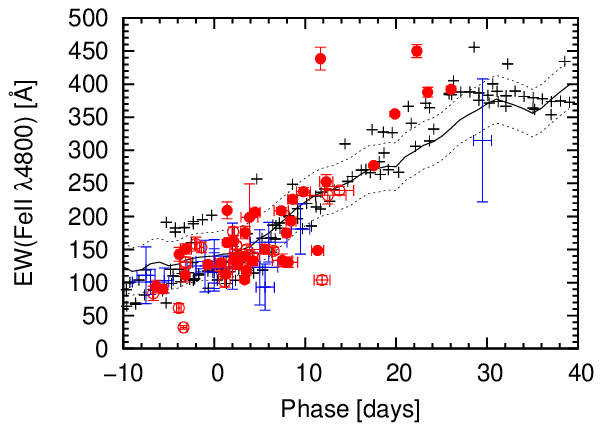} \\
\plottwo{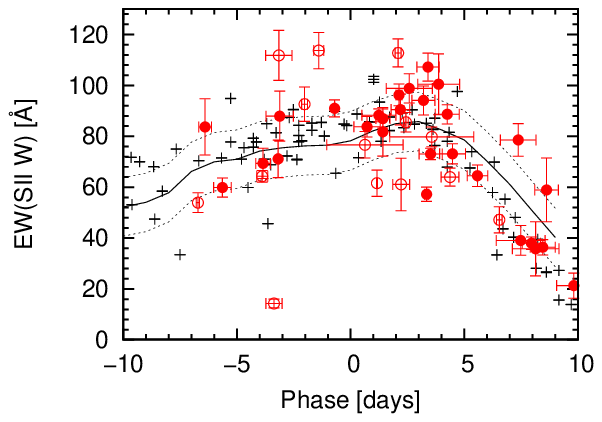}
{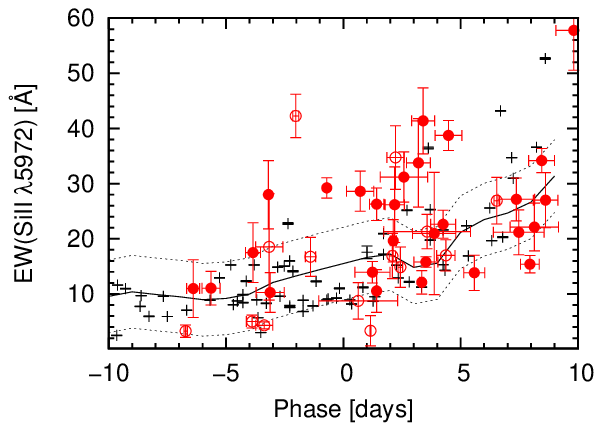} \\
\plottwo{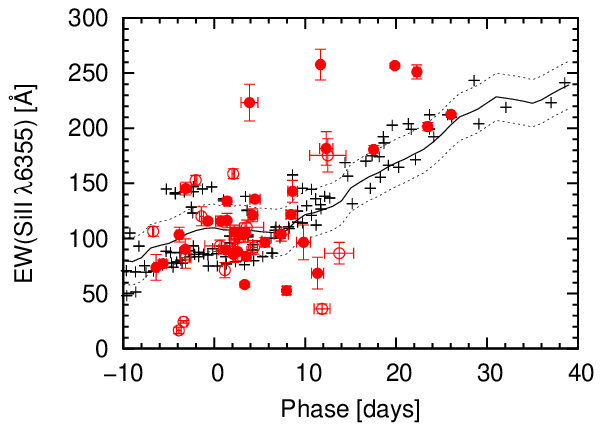}
{f7h.eps} \\
 \caption{The temporal evolution of equivalent widths for various elements. The symbols are the same as in Figure \ref{p_vel}. The SCP SNe Ia \citep{gar07} are marked in blue and the SNLS SNe Ia \citep{bro08} are marked in green. 
\label{p_ew}
  }
\end{figure}

In Figure \ref{p_ew}, we plot the temporal evolution of pEWs for all the measured lines of nearby and intermediate redshift SNe Ia (the same labels as Figure \ref{p_vel}): (a) ``{\ca}", (b) ``{\sia}", (c) ``{\mg}", (d) ``{\fe}", (e) ``{\sw}", (f) ``{\sib}" and  (g)``{\sic}".
One can see the following: 
(i) The ``{\ca}" pEW is 120 {\AA} at $p \sim 0$ and decreases by 50 {\AA}
to $p \sim 10$. 
(ii) The ``{\sia}" pEW stays nearly constant around 10 {\AA} at $p<0$ and
increases only gradually until $p \sim 7$. 
(iii) The ``{\mg}" pEW is around 100 {\AA} at $p<7$ and increases toward 
later phases.  Two SDSS SNe~Ia (SN12883 and SN12972) have a particulary 
small ``{\mg}" pEWs at $p \sim 10$. 
(iv) The ``{\fe}" pEW is around 100 {\AA} at $-5<p<0$ and it increases 
monotonically with SN phase to 400 {\AA} at $p \sim 30$.  {\FeII} is the 
element which dominates optical spectra as spectra evolve temporally 
\citep{bra05}.
(v) The ``{\sw}" pEW is large (80 {\AA}) until $p = 3$ and then begins to 
decrease. 
This feature is only visible until $p<10$.  \citet{bra05} explains that 
{\FeII} lines give a larger contribution to this feature at later phases.  
The temporal evolution for this feature has been displayed for the first time.
(vi) The ``{\sib}" pEW is 10 {\AA} at $p \sim -10$ and increases 
monotonically.  The ``{\sib}" feature is increasingly contaminated, again, 
mainly by NaI D, which dominates at $p \sim 10$.
(vii) The ``{\sic}" pEW is 90 {\AA} at $p \sim 0$ with the range of 30 {\AA}.  
It becomes large (200 {\AA}) at $p \sim 30$.  This is attributed to an 
iron feature \citep{bra08}.

Overplotted are high-z SNe Ia from \citet{blo06} (cyan pluses) and 
\citet{ell08} (green pluses) for velocities (Figure \ref{p_vel}).  
High-z SNe Ia from \citet{gar07} are in blue and those from \citet{bro08} 
in green for pEWs (Figure \ref{p_ew}). 
Since their spectra are not available to the public, we use values 
determined by these studies and averaged in the same manner as our sample. 
For the {\uvb} line, high-z SNe Ia from \citet{ell08} are in green pluses 
and theoretical predictions of \citet[]{len00} 
with the metallicities in the un-burned outer layer of 0.1, 0.3, 1.0 to 
3.0 solar are in gray lines (from top to bottom, Figure \ref{p_wvuv2}). 
The measurements for the \citet{hsi07} template are shown with a solid line.
The dispersions of the ``{\uvb}" wavelength for nearby and intermediate 
redshift SNe Ia (32 {\AA}) were larger than what can be explained by the 
variation of the progenitor metallicity.
Different sets of data are qualitatively compared in \S \ref{evo}.

\subsection{Redshift evolution \label{evo}}

\citet{bro08} examined the redshift dependences of the velocity for 
``{\ca}" and the pEWs for ``{\ca}", ``{\sia}" and ``{\mg}" within their 
high-z SNe Ia and showed that no systematic evolution of SN Ia properties 
were seen within the redshift ranges for SNLS SNe Ia.
\citet{fol08a} made ESSENCE SN Ia composite spectra for $0<z<0.6$ and 
measured ``{\ca}", ``{\sia}", and ``{\fe}" pEWs for these composites.  
They reported that the average ``{\fe}" pEW is lower for higher SNe Ia 
composites, possibly due to the larger number of bright and hot SNe~Ia in 
the high-z sample (but they also mention that this can not explain all 
the observed absorption profiles).
\citet{sul09} also constructed composites to report smaller average pEWs for 
``{\ca}", ``{\sia}" and ``{\mg}" with increasing redshifts. These effects 
are interpreted as the changing demographics of SNe Ia by these authors, 
that is a larger fraction of ``prompt \citep{man06}" SNe Ia at higher 
redshifts \citep{how07}. \citet{nor10} reported no signs of velocity differences 
in their sample and nearby counterparts, but found a fraction ($\sim 20\%$) 
of SNe Ia with lower pEWs than nearby averages, which was interpreted as partly 
caused by slightly peculiar SNe Ia (like e.g. the Shallow Silicon subtype, \citep{bra08}). 

We use all the {\gold} SNe Ia with a phase of $-5<p<5$, 
because their S/N are high and they are easy to compare with previous 
measurements.  The only exception is for ``{\siia}", for which we restrict the analysis
to the phase range $-5<p<3$, since this feature in the template spectra becomes 
unidentified at $p>3$.  
The ``{\siia}" feature can be seen in almost all spectra with the phase $-5<p<5$. 
The unidentification of this feature in the template might be attributed to a bias 
in the training spectra for the template.
In spite of several papers on the possibility of redshift 
evolution, few quantitative comparisons have been performed on a 
possible difference of distributions between samples of different 
redshift ranges for velocities and pEWs \citep{bro08}. 

\begin{figure}[htbp]
\plottwo{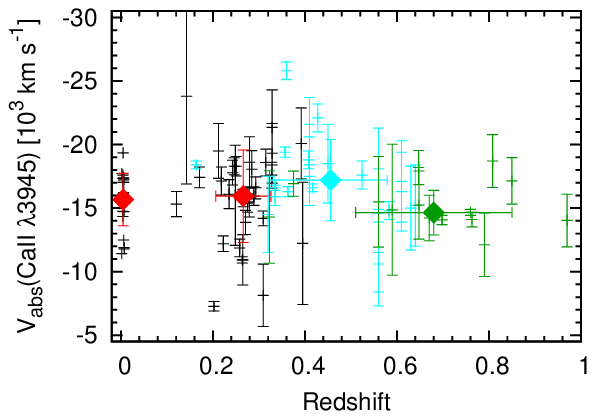}
{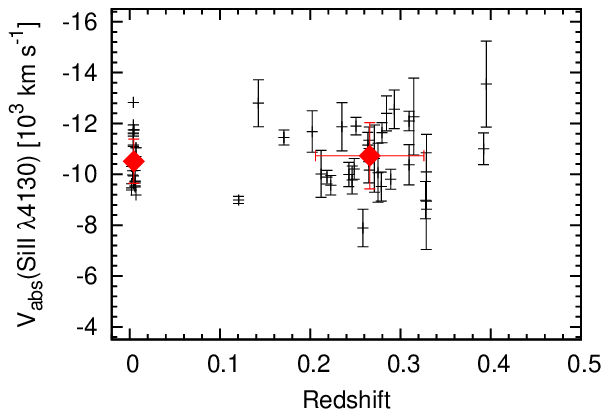} \\
\plottwo{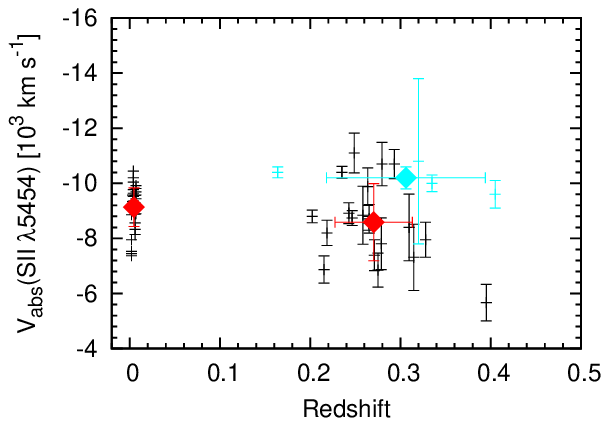}
{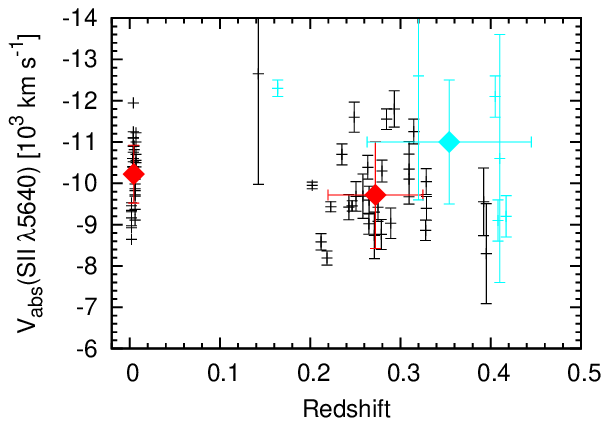} \\
\plottwo{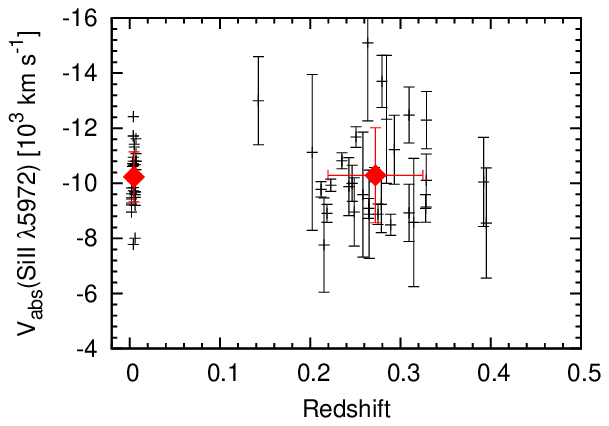}
{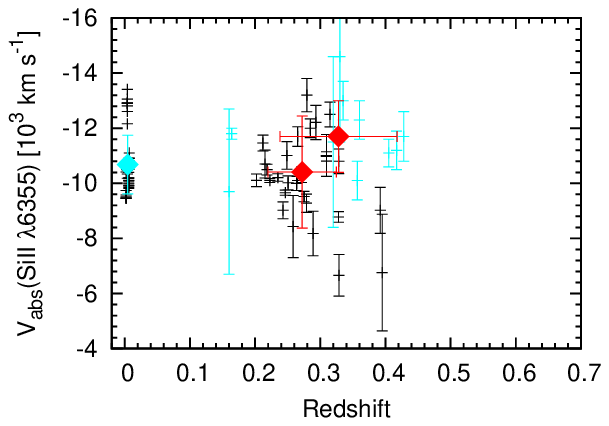} \\
\plottwo{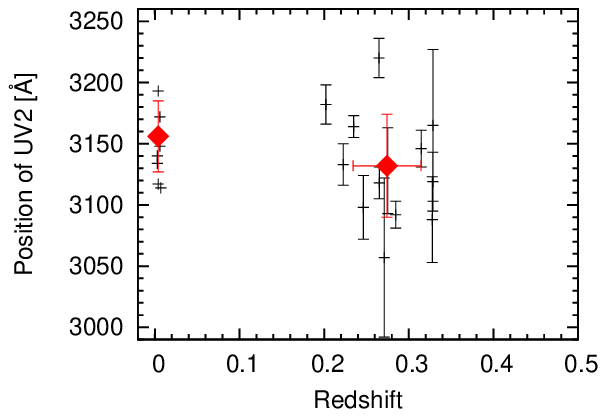}
{f8h.eps} \\
 \caption{Phase-corrected absorption line velocities as a function of redshift for various elements for nearby and intermediate redshift (black) and high-z SNe Ia (cyan and green). Two big red diamonds are averages and standard deviations of nearby and intermediate redshift samples. (see also text) \label{z_vel}
}
\end{figure}

\begin{figure}
\plottwo{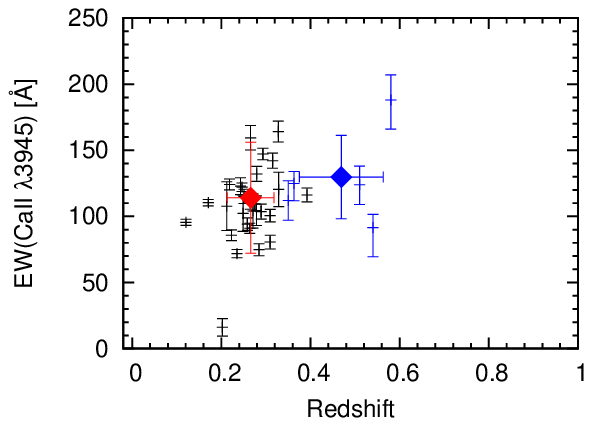}
{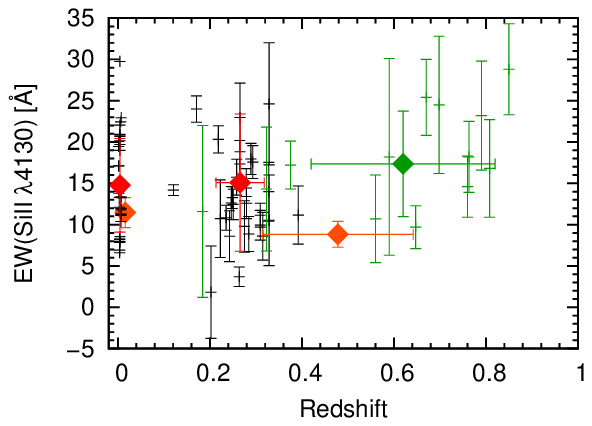} \\
\plottwo{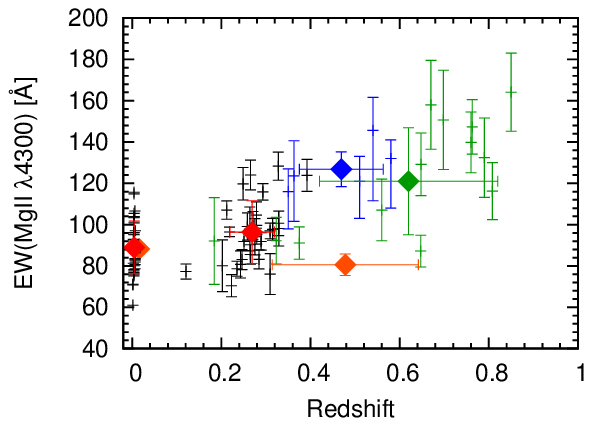}
{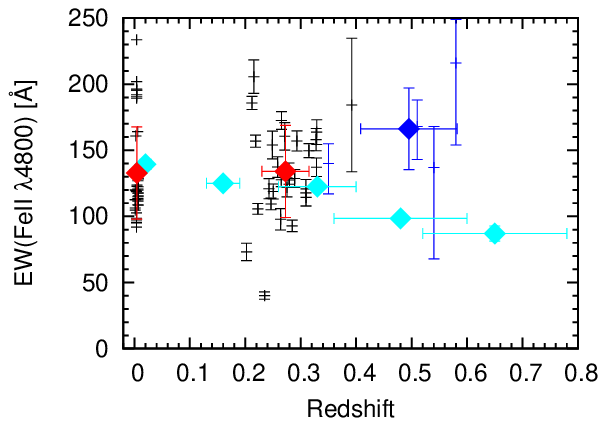} \\
\plottwo{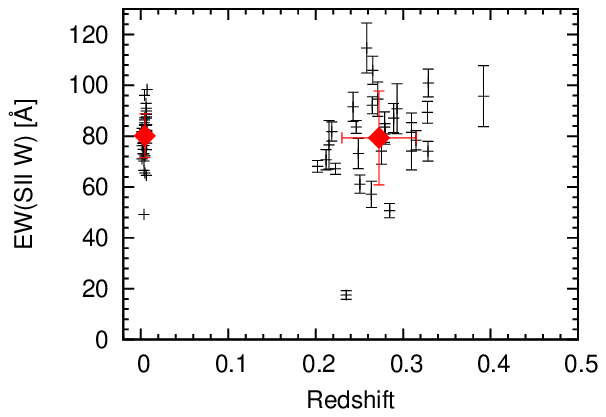}
{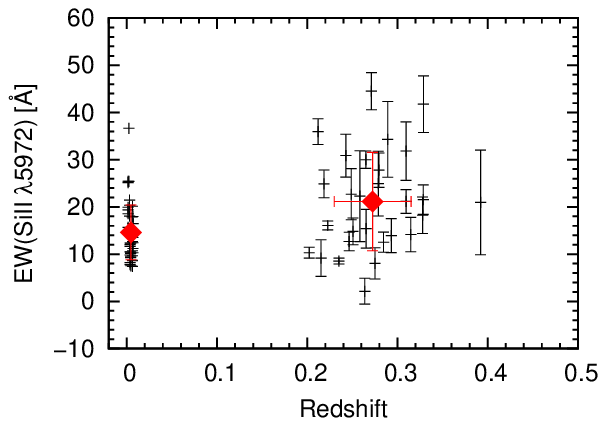} \\
\plottwo{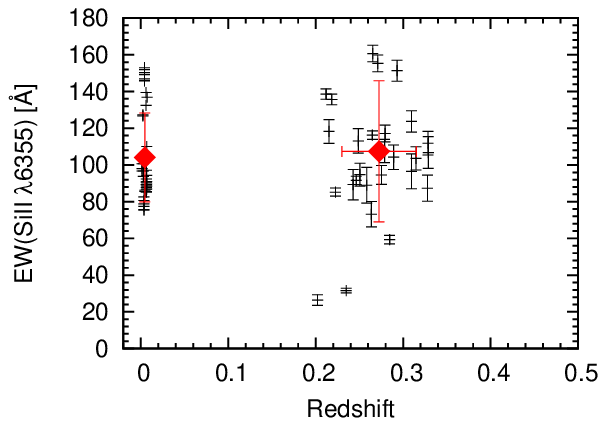}
{f9h.eps} \\
 \caption{Phase-corrected equivalent widths as a function of redshift for various elements for nearby and intermediate redshift (black) and high-z SNe Ia (cyan, blue, green). Two big red diamonds are averages and standard deviations of nearby and intermediate redshift samples. (see also text)
 \label{z_ew}
}
\end{figure}

An offset $E$ to the temporal evolution of the 
\citet{hsi07} template $X^{template}_p$ (and $1 \sigma$ deviation, 
$\sigma$) was estimated by expanding error bars on each trend until a 
fit to produce a {\rchi}$=\frac{1}{N-1} \sum (X_p-(\bar{X}^{template}_p+E))^2/(\sigma_p^2+\sigma^2$) of 1.0, where each SN measurement was 
($X_p$, $\sigma_p$).  $\sigma$ was assumed to be constant with phase so 
that the dispersions do not represent differences in observed numbers in 
each phase.  The temporal evolution of the \citet{hsi07} template shifted 
to represent the nearby SNe are shown as solid black lines in Figures 
\ref{p_vel} to \ref{p_ew}, with dotted lines for their 1$\sigma$ regions.
Figures \ref{z_vel} and \ref{z_ew} are velocities/wavelengths and pEWs as a 
function of redshift; ``{\ca}", ``{\sia}", ``{\siia}", ``{\siib}", ``{\sib}" 
and ``{\sic}" for velocities,``{\uvb}" for wavelengths and ``{\ca}", ``{\sia}", 
``{\mg}", ``{\fe}", ``{\sw}", ``{\sib}" and ``{\sic}" for pEWs. Overplotted 
are the measurements to individual high-z SNe Ia spectra of \citet{gar07} 
(blue pluses) and \citet{bro08} (green pluses) and composite spectra of 
\citet{fol08a} (cyan pluses) and \citet{sul09} (orange pluses).

Table \ref{velew_t} summarizes the nearby and intermediate redshift 
comparisons of velocities, and the ``{\uvb}" wavelengths and pEWs for 
each feature. 
Column 1 is the name of feature.  The number of SNe Ia, mean values and 
dispersions for nearby SNe Ia are listed from columns 2 to 4.  These 
parameters for intermediate redshift SNe Ia are from columns 5 to 7.  Column 8 is the 
probability of the Fisher F test that nearby and intermediate redshift SNe Ia have the 
same dispersion.  Column 9 is the probability of the Student t test that 
nearby and intermediate redshift SNe Ia have the same mean.  Since probabilities of the 
same dispersions vary, we calculate the Student's t value without 
assuming an equal variance between nearby and intermediate redshift samples. 
This table indicates that distributions for all the features of intermediate redshift SNe Ia are consistent with nearby SNe Ia counterparts.  The 
features for which different origins may be suspected are the ``{\siib}" 
velocity and the ``{\sib}" pEW.  We will discuss these suspected 
features in \S \ref{disc09b}.

\section{Discussion} \label{disc09b}
The temporal and redshift evolution of UV to optical features for intermediate redshift 
SNe Ia have been presented and compared with those of nearby SNe Ia 
counterparts. 
Statistical tests have been performed to investigate the possibility of 
a change in average properties (Table \ref{velew_t}).  All the features 
are consistent between nearby and intermediate redshift SNe Ia within 2$\sigma$ except 
for the ``{\siib}" velocity and the ``{\sib}" pEW.  This suggests that 
line profiles of intermediate redshift SNe Ia are similar to their nearby 
counterparts.  These findings support the picture that SNe Ia are a 
``standard" candle for the intermediate redshift range. 

We now examine the three features, ``{\siib}", ``{\sib}" and
``{\mg}", which have the lowest probability of being drawn
from identical populations in the low-z and intermediate redshift samples.
The ``{\siib}" feature is narrow and easily affected by sky noise.  Once 
an uncertainty in the mean value is taken into account, the nearby and 
intermediate redshift distributions become consistent at a significance level comparable
to the other features, the probability of belonging to the same distribution 
going up to 0.18. 
The redshift average of our sample is $z\sim 0.2$, at which the biggest 
atmospheric line tends to overlap the ``{\sib}" feature.  Although we 
attempted to remove sky lines completely, the larger deviation for the 
mid-z sample might be due to under/over subtraction of the unpredictable 
nature of the time variation. 
If, however, large pEWs for intermediate redshift SNe Ia are real, then SNe Ia with 
larger ``{\sib}" pEWs would be explained by cooler SNe Ia \citep{hac08}. 
The next lowest probability of belonging to the same distribution is for the 
``{\mg}" pEW: the average pEW of ``{\mg}" for intermediate redshift SNe Ia is 
$96{\AA}/89{\AA} \sim 8\%$ larger than the nearby average with a 12 \% 
probability.  \citet{tra05} simulated the metallicity effect on the 
detailed nucleosynthetic yields and showed that Mg is more synthesized 
by low metallicity progenitors. %as the $^{22}$Ne mass goes down and the $^{12}$C mass goes up in SN Ia progenitors, while the other elements are unchanged.
Assuming that the Mg synthesis rate is proportional to the ``{\mg}" pEW, 
an $\sim 8$ \% mass increase corresponds to a decrease of progenitor 
metallicity by $\sim$0.2 {\Zsun} \citep{tra05}.  This is comparable to 
the average decrease of metallicity between nearby and intermediate redshift galaxies 
\citep{rod08}.

We found similar velocity distributions for all the 
features of intermediate redshift and nearby SNe Ia (Figure \ref{p_vel} and 
Table \ref{velew_t}).  This finding is consistent with \citet{nor10}.  
However, we do not find significant differences in pEWs of ``{\sia}" 
(called ``f2" in their paper) and ``{\fe}" (called ``f4" in their paper).  
The intermediate redshift 
sample of \citet{nor10} contains a larger fraction of SNe Ia with high 
stretch (\citet{guy05}, equivalent to {\xx}) than their nearby sample.  
Note that SNe Ia with high stretch have larger ``{\sia}" pEWs 
\citep{ars08,nor10}.  
They also noticed a pEW-deficit subsample containing several SNe that likely 
would have been identified as peculiar if observed locally with higher S/N, 
possibly explaining the systematic pEW differences.
Our intermediate redshift sample has the average {\xx} of $-0.06 \pm 1.13$ for $-5<p<5$ 
and this value is comparable to our nearby sample ($-0.10 \pm 0.48$) in 
the same phase range.  Thus, the inconsistency for this feature can be 
explained by the sample difference. 

Some features for intermediate redshift SNe Ia can be compared with high-z trends from 
previous studies \citep{blo06,gar07,bro08,sul09} in Figures \ref{z_vel} 
and \ref{z_ew}.  Since their spectra are not available to the public, we 
use values determined in their papers and treated them in the same manner 
as we did our sample.
For each feature, a high-z sample used in a previous study with the 
largest dataset was compared with intermediate redshift SNe Ia. We found in general 
the high-z and intermediate redshift samples to be similar.  The velocity distributions for 
``{\ca}", ``{\siia}", ``{\siib}", and ``{\sic}" come from the same 
population with $>19$ \% probability.  The pEW distributions agree for 
``{\ca}", "{\sia}", and ``{\fe}" with $>16$ \% probability.  Also the 
mean ``{\fe}" pEW at $z=0.33$ \citep{fol08a} is consistent with our 
distribution. 

\begin{figure}
\plotone{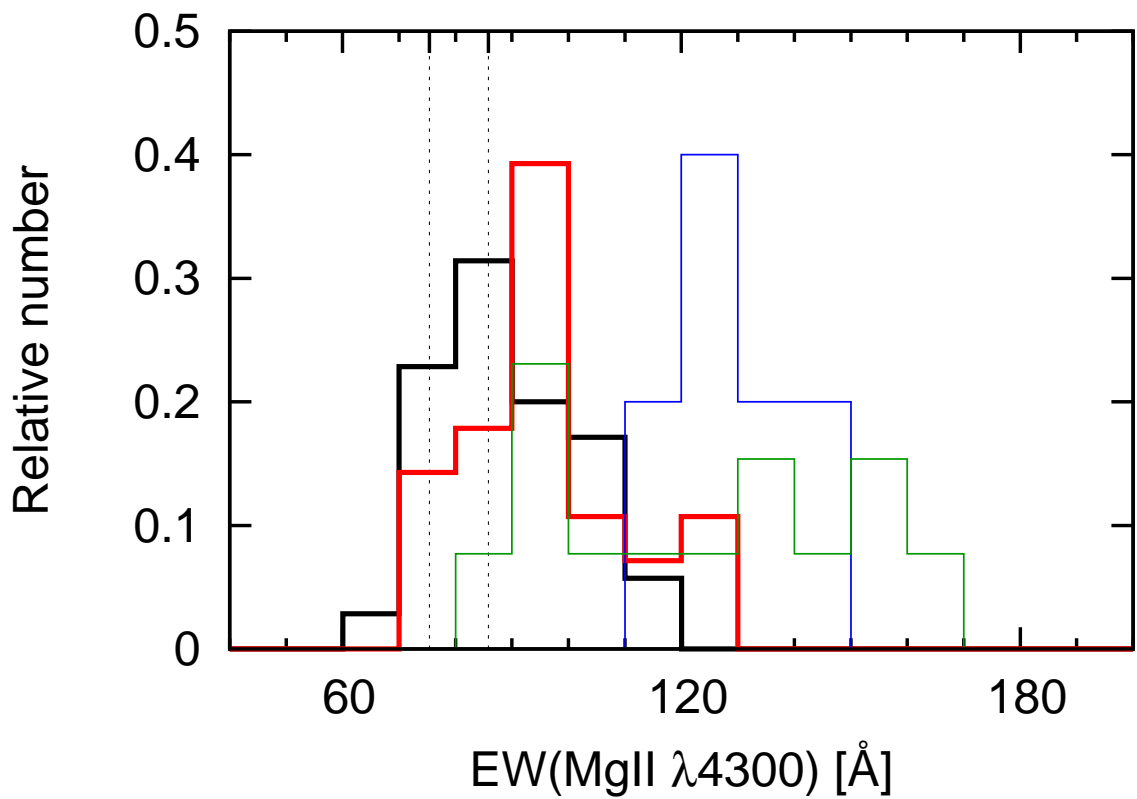}
\caption{Histograms of phase-corrected equivalent widths for the ``{\mg}" feature: nearby (black) and intermediate redshift sample (red). The 1$\sigma$ pEW region for the \citet{sul09} high-z sample is shown by two black-dotted lines. Blue and green histograms are for the \citet{gar07} and \citet{bro08} high-z samples. \label{z_ew_mg}}
\end{figure}

However, the behavior of the ``{\mg}" pEW is hard to 
understand.  Figure \ref{z_ew_mg} is the histogram of the ``{\mg}" pEW 
distribution along with previous measurements.  Distributions for nearby 
and intermediate redshift SNe Ia are shown in black and red.  The 1$\sigma$ pEW region 
for the \citet{sul09} high-z sample is shown by two black-dotted lines.  
Blue and green histograms are for high-z SNe Ia by \citet{gar07} and 
\citet{bro08}.  High-z distributions differ from intermediate redshift by more than 
$3\sigma$: the pEW for the composite spectrum of \citet{sul09} 
($\bar{z}=0.48$) is lower than the intermediate redshift average.  They interpreted this 
as a larger fraction of hotter SNe Ia in the high-z Universe.  This 
interpretation explains the larger fraction of SNe Ia with wide 
lightcurves \citep{how07}, however it does not explain the line profile 
of the ``{\fe}" feature for high-z SNe Ia \citep{fol08a}.  The average 
pEWs for the \citet{gar07} or \citet{bro08} samples are, on the other 
hand, higher than the intermediate redshift average.  Due to the finding of \citet{how07}, 
lower ``{\mg}" pEW can be interpreted as a difference in the progenitor 
abundance \citep{tra05} rather than as a larger fraction of cooler SNe Ia.

A spectroscopic luminosity indicator calibrates the maximum luminosity from 
a spectroscopic observable as does a lightcurve width \citep[e.g.][]{phi93} 
or a color \citep[e.g.][]{tri98}. 
A flux ratio and pEWs have been suggested as its candidate. \citet{bai09} 
presented that the flux ratio is best correlated with maximum absolute magnitude 
is $F(6420{\AA})/F(4430{\AA})$.
Several papers have introduced an pEW of the ``{\sia}" feature \citep{bro08}, 
the ``{\sib}" feature \citep{hac06} and the ratio of the ``{\sib}" pEW to 
the {\sic}" pEW \citep{hac06} as a spectroscopic luminosity indicator. 
Suitable indicators should have a large intrinsic dispersion and small amount of 
pEW  change with phase. As has been shown in Figure \ref{p_ew}, the ``{\sia}" 
pEW shows the highest ratio of the pEW dispersion around the mean, and the least 
temporal evolution during $-5<p<+5$. 
Moreover, the ``{\sia}" pEW is redshifted to wavelength regions less affected 
by night sky at intermediate redshift and can be also observed in the optical 
spectrograph for SNe Ia at $z \lesssim 1.2$. It can be said that the ``{\sia}" 
pEW is the best suited as a luminosity indicator among optical {\SiII} lines. 
An indication of maximum luminosity from pEWs will be discussed in \citet{kon09c}.

We measured line velocities and pEWs of nearby spectra as well as intermediate 
redshift spectra, but, we had to use values of line velocities and pEWs 
presented in papers for high-z SNe Ia. 
Since it is difficult to remove the contribution of the host galaxy's light 
completely at high-z, an unknown systematic error might remain in measurements 
by different authors. 
An uniform method for galaxy subtraction and feature measurements would be 
desired to compare spectra in different redshifts, after spectra have been 
reduced by various methods optimized by their telescopes.

\section{Conclusion} \label{sum09b}

We have derived line velocities, peak wavelengths and pEWs 
of spectral features during $-10<p<40$ for 59 normal intermediate redshift SNe Ia.  
Additionally, we compared the line profiles around the date of maximum 
brightness with nearby SNe Ia counterparts.  We conclude the following: 
line velocities and pEWs for intermediate redshift SNe Ia are consistent with 
their nearby counterparts within $2\sigma$.  This supports the picture 
that SNe Ia are a ``standard" candle for the intermediate redshift range as has been 
shown between SNe Ia at nearby and high redshifts.  Although the 
probability that the velocity of the ``{\siib}" feature for intermediate redshift and nearby 
SNe Ia comes from the same population shows a low probability, these 
samples are consistent within their uncertainties.  A hint of larger ``{\mg}" 
pEW distribution for intermediate redshift than nearby SNe Ia could be interpreted as 
the difference in the progenitor abundance.

\acknowledgements
Acknowledgements -- 
We thank Harold Spinka for his careful reading of the manuscript for publication.
K.K. thanks the COE Program "the Quantum Extreme Systems and Their
Symmetries" for fiscal 2007, the Global COE Program "the Physical
Sciences Frontier" for fiscal 2008, MEXT, Japan and the JASSO
scholarship for fiscal 2007-2009. L.{\"{O}} is partially supported 
by the Spanish Ministry of Science and Innovation (MICINN) through 
the Consolider Ingenio-2010 program, under project CSD2007-00060 
``Physics of the Accelerating Universe (PAU)''.

Funding for the SDSS and SDSS-II was provided by the Alfred P. Sloan
Foundation, the Participating Institutions, the National Science
Foundation, the U.S. Department of Energy, the National Aeronautics and
Space Administration, the Japanese Monbukagakusho, the Max Planck
Society, and the Higher Education Funding Council for England. 
\url{The SDSS Web site is http://www.sdss.org/}.

The SDSS is managed by the Astrophysical Research Consortium (ARC) for
the Participating Institutions. The Participating Institutions are The
University of Chicago, Fermilab, the Institute for Advanced Study, the
Japan Participation Group, The Johns Hopkins University, Los Alamos
National Laboratory, the Max-Planck-Institute for Astronomy (MPIA), the
Max-Planck-Institute for Astrophysics (MPA), New Mexico State
University, University of Pittsburgh, Princeton University, the United
States Naval Observatory, and the University of Washington.  
This research has made use of the NASA/IPAC Extragalactic Database (NED)
which is operated by the Jet Propulsion Laboratory, California Institute
of Technology, under contract with the National Aeronautics and Space
Administration. 
Facilities: \facility{SDSS}, \facility{Subaru(FOCAS)}

\clearpage

\clearpage

\clearpage

\clearpage

\clearpage

\clearpage

\clearpage

\begin{deluxetable}{llrrllr} 
 \tabletypesize{\scriptsize} 
 \tablecaption{Normal Type Ia Supernovae observed by Subaru/FOCAS \label{sbr1aspecinfo}} 
 \tablewidth{0pt} 
% \rotate 
 \tablehead{ 
  \colhead{SDSS ID} & \colhead{IAU name\tablenotemark{a}} &
  \colhead{$E(B-V)_{MW}$ \tablenotemark{b}} & 
  \colhead{MJD \tablenotemark{c}} & \colhead{Redshift \tablenotemark{d}} &
  \colhead{Sample \tablenotemark{e}} & \colhead{SN phase \tablenotemark{f}}}
 \startdata 
 1119 &  2005fc &  0.060 &  53640.26 &  0.2974 (ge) &  PS & 12.5 \\ 
 1166 &  -- &  0.019 &  53641.39 &  0.3814 (ga) &  G & 7.5 \\ 
 1686 &  -- &  0.055 &  53640.42 &  0.1362 (ge) &  G & 11.7 \\ 
 1688 &  -- &  0.050 &  53641.25 &  0.3591 (ge) &  G & 8.1 \\ 
 2165 &  2005fr &  0.033 &  53641.42 &  0.2841 (sn) &  G & 5.6 \\ 
 2330 &  2005fp &  0.024 &  53641.35 &  0.2129 (ge) &  G & 4.5 \\ 
 2422 &  2005fi &  0.039 &  53641.32 &  0.2628 (ge) &  SS & 1.2 \\ 
 2635 &  2005fw &  0.096 &  53670.60 &  0.1434 (ge) &  G & 23.5 \\ 
 2789 &  2005fx &  0.051 &  53640.38 &  0.2862 (ga) &  G & 3.2 \\ 
 2992 &  2005gp &  0.087 &  53669.63 &  0.1261 (ge) &  G & 22.3 \\ 
 3080 &  2005ga &  0.048 &  53670.36 &  0.1743 (ge) &  G & 17.5 \\ 
 3451 &  2005gf &  0.056 &  53640.31 &  0.2497 (ge) &  G & -3.8 \\ 
 3452 &  2005gg &  0.059 &  53640.34 &  0.2311 (ge) &  G & -6.4 \\ 
 5391 &  2005hs &  0.111 &  53669.56 &  0.3003 (ge) &  G & 8.6 \\ 
 5533 &  2005hu &  0.070 &  53669.25 &  0.2199 (ge) &  G & 1.4 \\ 
 5717 &  2005ia	&  0.031 &  53670.28 &  0.2521 (ge) &  SS & 4.4 \\ 
 5737 &  2005ib &  0.038 &  53670.32 &  0.3933 (ge) &  G & 3.9 \\ 
 5751 &  2005hz &  0.022 &  53701.33 &  0.1308 (ge) &  SS & 27.1 \\ 
 5844 &  2005ic &  0.121 &  53669.29 &  0.3120 (ge) &  G & 4.3 \\ 
 5944 &  2005hc &  0.030 &  53701.49 &  0.0445 (ge) &  G & 26.0 \\ 
 5957 &  2005ie &  0.036 &  53669.60 &  0.2796 (ge) &  G & 2.2 \\ 
 6057 &  2005if &  0.119 &  53700.54 &  0.0669 (ge) &  SS & 28.5 \\ 
 6108 &  2005ih &  0.054 &  53671.33 &  0.2599 (ge) &  SS & -3.2 \\ 
 6196 &  2005ig &  0.063 &  53670.25 &  0.2811 (ga) &  G & 0.7 \\ 
 6249 &  2005ii &  0.041 &  53671.39 &  0.2947 (ge) &  G & -3.1 \\ 
 6406 &  2005ij &  0.081 &  53700.50 &  0.1244 (ge) &  G & 19.9 \\ 
 6699 &  2005ik &  0.053 &  53671.29 &  0.3105 (ge) &  G & -3.2 \\ 
 9032 &  2005le &  0.060 &  53700.23 &  0.2541 (ge) &  G & 8.5 \\ 
 9207 &  2005lg &  0.038 &  53700.47 &  0.3496 (ge) &  G & 9.8 \\ 
 10449 &  2005ll &  0.062 &  53701.23 &  0.2415 (ge) &  G & 2.6 \\ 
 10550 &  2005lf &  0.036 &  53700.30 &  0.3000 (ge) &  G & 7.4 \\ 
 12855 &  2006fk &  0.052 &  53993.27 &  0.1722 (ga) &  G & -2.2 \\ 
 12860 &  2006fc &  0.069 &  53993.28 &  0.1210 (ge) &  G & -0.3 \\ 
 12869 &  2006ge &  0.140 &  53996.24 &  0.2841 (ge) &  PS & 13.7 \\ 
 12883 &  2006fr &  0.102 &  53995.30 &  0.3060 (ge) &  SS & 11.9 \\ 
 12972 &  2006ft &  0.023 &  53995.43 &  0.2606 (ge) &  G & 11.4 \\ 
 12977 &  2006gh &  0.025 &  53996.50 &  0.2474 (ge) &  G & 1.2 \\ 
 13025 &  2006fx &  0.083 &  53995.35 &  0.2240 (ge) &  G & 3.5 \\ 
 13099 &  2006gb &  0.039 &  53995.39 &  0.2655 (ge) &  G & -0.7 \\ 
 13152 &  2006gg &  0.021 &  53996.40 &  0.2031 (ge) &  SS & -3.9 \\ 
 13174 &  2006ga &  0.026 &  53995.48 &  0.2357 (ge) &  SS & -3.4 \\ 
 13327 &  2006jf &  0.057 &  53996.32 &  0.2814 (ge) &  PS & -21.5 \\ 
 13934 &  2006jg &  0.102 &  54023.41 &  0.3310 (ge) &  G & 12.3 \\ 
 14261 &  2006jk &  0.094 &  54023.27 &  0.2853 (ge) &  G & 3.3 \\ 
 14298 &  2006jj &  0.076 &  54023.23 &  0.2692 (sn) &  G & 8.0 \\ 
 14456 &  2006jm &  0.090 &  54023.46 &  0.3283 (ge) &  G & 3.4 \\ 
 15217 &  2006jv &  0.029 &  54023.49 &  0.371 (sn) &  PS & -1.4 \\ 
 15219 &  2006ka &  0.037 &  54023.54 &  0.2467 (ge) &  G & -5.6 \\ 
 16631 &  2006pv &  0.128 &  54065.53 &  0.207 (sn) &  G & 1.4 \\ 
 16758 &  2006pw &  0.085 &  54065.25 &  0.3261 (ge) &  PS & 2.4 \\ 
 16776 &  2006qd &  0.044 &  54066.30 &  0.2667 (ge) &  PS & 2.1 \\ 
 16779 &  2006qa &  0.105 &  54065.44 &  0.3983 (ge) &  SS & 6.5 \\ 
 16781 &  2006qb &  0.111 &  54065.48 &  0.3257 (ge) &  PS & 3.6 \\ 
 16847 &  2006px &  0.033 &  54066.30 &  0.2767 (ge) &  PS & 0.6 \\ 
 16938 &  2006qe &  0.047 &  54065.29 &  0.386 (sn) &  S & 2.2 \\ 
 16953 &  2006pp &  0.024 &  54065.33 &  0.3387 (ge) &  G & 2.1 \\ 
 17048 &  2006qi &  0.025 &  54065.36 &  0.189 (sn) &  S & -6.7 \\ 
 17081 &  2006ql &  0.038 &  54065.40 &  0.2749 (ga) &  PS & -2.0 \\ 
 17117 &  2006qm &  0.032 &  54066.38 &  0.1404 (ge) &  PS & -12.6 \\ 
 \enddata 
%\tablecomments{RA and Dec in Columns 4 and 5 are the Right Ascension and Declination (J2000) in degrees.}
\tablenotetext{a}{IAU names were not attached to three SNe due to on-site analysis.}
\tablenotetext{b}{Color excess within our Galaxy.}
\tablenotetext{c}{The observational mid-date. The red part of a spectrum is first observed and the blue part follows.}
\tablenotetext{d}{Heliocentric redshift. ``ge'' indicates that the redshift of the target was determined from galaxy emission line(s), ``ga'' from galaxy absorption line(s) and ``sn'' for redshift from SN spectrum fitting.}
\tablenotetext{e}{The validity of the data; 
G: {\it gold} sample,  PS: {\it silver} lightcurves, 
SS: {\it silver} spectrum,  S: {\it silver} lightcurves \& spectrum}
\tablenotetext{f}{The SN phase is the rest-frame phase in days from their maximum brightness date.}
\end{deluxetable}

\begin{deluxetable}{lrrrr} 
 \tabletypesize{\scriptsize} 
 \tablecaption{
 Type Ia Supernovae in the {\bf Opt} sample
 \label{optspecinfo}
 } 
 \tablewidth{0pt} 
% \rotate 
 \tablehead{ 
 \colhead{IAU Name} &  \colhead{RA} &  \colhead{Dec} &
 \colhead{$E(B-V)_{MW}$ \tablenotemark{a}} & \colhead{$z_{spec}$}
 }
 \startdata 
 1981B  &  188.62 &   2.20 &  0.013 & 0.0060 \\ 
 1989B  &  170.06 &  13.01 &  0.013 & 0.0024 \\ 
 1990N  &  190.74 &  13.26 &  0.018 & 0.0034 \\ 
 1991M  &  239.65 &  17.46 &  0.030 & 0.0072 \\ 
 1994D  &  187.87 &   7.98 &  0.013 & 0.0015 \\ 
 1996X  &  199.50 & -26.85 &  0.068 & 0.0069 \\ 
 1998aq &  179.11 &  55.13 &  0.014 & 0.0037 \\ 
 1998bu &  169.69 &  11.84 &  0.025 & 0.0030 \\ 
 1999ee &  334.04 & -36.84 &  0.059 & 0.0114 \\ 
 2000E  &  309.31 &  66.10 &  0.371 & 0.0047 \\ 
 2002bo &  154.53 &  21.83 &  0.051 & 0.0042 \\ 
 2003du &  218.65 &  59.33 &  0.010 & 0.0064 \\ 
 \enddata 
\tablecomments{RA and Dec in Column 2 and 3 are the Right Ascension and Declination (J2000) in degrees.}
% \tablecomments{TABLE COMMENTS} 
 \tablenotetext{a}{Color excess within our Galaxy.}
\end{deluxetable} 

\begin{deluxetable}{lrrrr} 
 \tabletypesize{\scriptsize} 
 \tablecaption{
 Type Ia Supernovae in the {\bf UV} sample
 \label{uvspecinfo}
 } 
 \tablewidth{0pt} 
% \rotate 
 \tablehead{ 
 \colhead{IAU Name} &  \colhead{RA} &  \colhead{Dec} &
 \colhead{$E(B-V)_{MW}$ \tablenotemark{a}} & \colhead{$z_{spec}$}
}
 \startdata 
 1980N  &   50.75 &  -37.21 &  0.046 & 0.0059 \\ 
 1981B  &  188.62 &    2.20 &  0.013 & 0.0060 \\ 
 1982B  &  108.64 &   84.39 &  0.064 & 0.0074 \\ 
 1983G  &  193.09 &   -1.20 &  0.010 & 0.0041 \\ 
%1986G  &  201.40 &  -43.03 &  0.187 & 0.0018 \\ 
 1989B  &  170.06 &   13.01 &  0.013 & 0.0024 \\ 
 1990N  &  190.74 &   13.26 &  0.018 & 0.0034 \\ 
%1991bg &  186.27 &   12.87 &  0.012 & 0.0035 \\ 
 2001ba &  174.51 &  -32.33 &  0.060 & 0.0296 \\ 
 2001el &   56.13 &  -44.64 &  0.094 & 0.0039 \\ 
 \enddata 
\tablecomments{RA and Dec in Column 2 and 3 are the Right Ascension and Declination (J2000) in degrees.}
% \tablecomments{TABLE COMMENTS} 
 \tablenotetext{a}{Color excess within our Galaxy.}
\end{deluxetable} 

\begin{deluxetable}{cllll}
 \tabletypesize{\scriptsize}
 \tablecaption{Line measurements\label{feature}}
 \tablewidth{0pt}
 \tablehead{
 \colhead{Feature ID} & \colhead{Label} & 
 \colhead{Wavelength ranges$^a$} & 
 \colhead{} &
 \colhead{Rest Wavelength [\AA]$^b$} \\
 \colhead{} & \colhead{} & 
 \colhead{Absorption minimum} & 
 \colhead{Absorption boundaries (blue, red)} &
 \colhead{}
 }
 \startdata
 UV2 & ``{\uvb}'' & 3050:3200 & -, -                  & - \\
 1   & ``{\ca}''  & 3650(3700):4000 & 3500(4000$^c$):3720, 3860:4050 & 3945 \\
 2   & ``{\sia}'' & 3950:4100 & 3860:4000, 4000:4220 & 4130 \\
 3   & ``{\mg}''  & - & 3900(4000$^d$):4150, 4450:4700 & - \\
 4   & ``{\fe}''  & - & 4450:4700, 5050:5590 & - \\
 5   & ``{\sw}''  & 5240:5365, 5400:5545 & 5080:5400, 5500:5700 & 5454,5640 \\
 6   & ``{\sib}'' & 5650:5850 & 5550:5700, 5800:6000 & 5972 \\
 7   & ``{\sic}'' & 5900:6300 & 5820:6000, 6200:6540(6400$^e$) & 6355 \\
 \enddata
 \tablenotetext{a}{We select different search wavelength ranges from \citet{gar07} to optimize for automated measurements of our dataset. A wavelength of an absorption minimum or a blue/red absorption boundary is measured within the range of $\lambda_a$ and $\lambda_b$ for $\lambda_a:\lambda_b$.}
 \tablenotetext{b}{The rest-frame wavelengths for the transition of the given ion; 
 the weighted mean wavelength of the two strongest transitions for ID1 and ID7,
 and the wavelength of the strongest transition for ID2, ID5 and ID6 \citep{blo06}.}
 \tablenotetext{c}{For the phase range of $-10.0 \leq p < +0.0$.}
 \tablenotetext{d}{For the phase range of $-10.0 \leq p < +6.5$. The ``{\sia}'' feature gets merged to ``{\mg}" at $p>6.5$.}
 \tablenotetext{e}{For the phase range of $-10.0 \leq p < +5.0$.} 
\end{deluxetable}

\begin{deluxetable}{lrrrrrrrr}
 \tabletypesize{\scriptsize}
 \tablecaption{
 The nearby and intermediate redshift comparisons for each feature \label{velew_t}
 }
 \tablewidth{0pt}
 \tablehead{
 \colhead{Features} & \colhead{Nearby} & \colhead{} & \colhead{} & 
 \colhead{Mid-z} & \colhead{} & \colhead{} & 
 \colhead{P(F test)$^a$} & \colhead{P(T test)$^b$} \\
 \colhead{} & \colhead{N} & \colhead{mean} & \colhead{deviation} & 
 \colhead{N} & \colhead{mean} & \colhead{deviation} &
 \colhead{} & \colhead{}
 }
 \startdata
 Velocity$^c$ \\
 {\ca}   & 23 & -15.7 & 21.0 & 19 & -16.2 & 23.5 & 0.61 & 0.44     \\
 {\sia}  & 34 & -10.5 &  8.6 & 19 & -10.5 &  9.2 & 0.72 & 0.79     \\
 {\siia} & 38 & -9.1  &  6.9 & 11 & -8.9  & 11.7 & 0.02 & 0.53     \\
 {\siib} & 46 & -10.2 &  6.8 & 18 & -9.6  & 11.0 & 0.02 & 0.05$^e$ \\
 {\sib}  & 45 & -10.2 &  8.9 & 18 & -10.0 & 11.9 & 0.12 & 0.39     \\
 {\sic}  & 47 & -10.7 & 10.6 & 18 & -10.3 & 14.4 & 0.10 & 0.26     \\
 \hline\hline
 Wavelength $^d$ \\
 {\uvb} & 8 & 3146 & 32 & 6 & 3147 & 32 & 0.92 & 0.18 \\
 \hline\hline
 Equivalent width $^d$ \\
  {\ca}  & 16 & 103 &  23 &  19 & 112 &  24 & 0.88 & 0.27    \\
 {\sia} & 35 &   15 &   6 &  19 &  16 &   6 & 0.97 & 0.56    \\
 {\mg}  & 35 &   89 &  13 &  18 &  96 &  16 & 0.30 & 0.12    \\
 {\fe}  & 44 &  133 &  24 &  18 & 138 &  24 & 0.95 & 0.46    \\
 {\sw}  & 47 &   80 &   9 &  18 &  81 &  11 & 0.28 & 0.73    \\
 {\sib} & 46 &   15 &   6 &  18 &  23 &   8 & 0.13 & $<$0.01 \\
 {\sic} & 46 &  104 &  24 &  18 & 113 &  24 & 0.95 & 0.19    \\
\enddata
 \tablenotetext{a}{A probability that nearby and intermediate redshift SNe Ia have the same dispersion.}
 \tablenotetext{b}{A probability that nearby and intermediate redshift SNe Ia have the same mean.}
 \tablenotetext{c}{Values in column 3, 4, 6, 7 are velocities in units of \vel{$10^3$}. The mean values in columns 3 and 6 are offsets from the trends of the \citet{hsi07} template spectra.}
 \tablenotetext{d}{Values in column 3, 4, 6, 7 are wavelengths or EWs in units of {\AA}. The mean values in columns 3 and 6 are offsets from the trends of the \citet{hsi07} template spectra.}
 \tablenotetext{e}{The probability from the same distribution going up to 0.18, when an uncertainty on the mean value is taken into account.}
\end{deluxetable}

\clearpage

\end{document}